\pdfoutput=1
\documentclass[letterpaper]{jpc}

\usepackage[T1]{fontenc}
\usepackage{lmodern}
\usepackage{graphicx}
\usepackage{booktabs}
\usepackage[authoryear,round]{natbib}
\usepackage{xurl}
\allowdisplaybreaks[2]
\hypersetup{
  pdftitle={Reporting Differential Privacy Guarantees via Operationally Equivalent Epsilon},
  pdfauthor={Masayuki Terada, Shinsuke Ito, Takumasa Ishioka, and Hidetoshi Matsui}
}
\makeatletter
\let\endfront@text\@empty
\let\enddoc@text\@empty
\makeatother

\begin{document}

\title[Reporting Privacy Guarantees via Operationally Equivalent $\varepsilon$]{Reporting Differential Privacy Guarantees via Operationally Equivalent $\varepsilon$:
What Does $\varepsilon=19.61$ Mean for the 2020 U.S. Census?\rsuper*}
\titlecomment{{\lsuper*}This paper revises and extends a preliminary Japanese version published in the proceedings of the Computer Security Symposium 2026 (CSS 2026) \citep{TI26-en}.}
\author[M. Terada]{Masayuki Terada}
\address{Faculty of Engineering, Kyoto Tachibana University}
\author[S. Ito]{Shinsuke Ito}
\address{Faculty of Economics, Chuo University}
\author[T. Ishioka]{Takumasa Ishioka}
\address{Faculty of Engineering, Kyoto Tachibana University}
\author[H. Matsui]{Hidetoshi Matsui}
\address{Faculty of Data Science, Shiga University; Center for Informatics Research and Education, Kyoto Tachibana University}

\begin{abstract}
The large value $\varepsilon=19.61$, reported as an approximate differential privacy (ADP) parameter for the 2020 U.S. Census, has prompted doubts about the adequacy of Census privacy protection. This value was obtained by converting component guarantees expressed in zero-concentrated differential privacy (zCDP), an extension of differential privacy (DP), to ADP at $\delta=10^{-10}$. Summarizing a guarantee by a single $(\varepsilon,\delta)$ pair can substantially loosen its upper bound on the power (true positive rate) of a distinguishing test at a given significance level. We express the power bound from the original guarantee as the $\alpha$-operationally equivalent $\varepsilon$: the smallest pure DP parameter yielding the same power bound at significance level $\alpha$. The full operationally equivalent $\varepsilon$ curve represents the original bound exactly on $(0,1)$, even when that bound is nonconcave. We also characterize the extrema of operationally equivalent $\varepsilon$ over an interval of significance levels and interpret them, when finite, as the parameters of the pure DP power bounds that enclose the original power bound on that interval. The mapping preserves the ordering of guarantees by their power bounds at each significance level, whereas conversion to single ADP pairs or Gaussian DP (GDP) envelopes can reverse this ordering for Laplace and Gaussian mechanisms with equal noise variance. For the P.L.~94-171 redistricting data, we reproduce the reported $\varepsilon=19.61$ as the sum of rounded component ADP values. The published zCDP guarantee yields operationally equivalent $\varepsilon$ values of approximately $4.25$, $2.94$, and $3.18$ at significance levels of $1\%$, $5\%$, and $10\%$, respectively. These values convey the original guarantee's constraints on distinguishing attacks more clearly than does the reported value $\varepsilon=19.61$, which by itself places little restriction on the success of such attacks.
\end{abstract}

\maketitle

\section{Introduction}\label{sec:introduction}

The U.S. Census Bureau adopted zero-concentrated differential privacy (zCDP), an extension of differential privacy (DP), for the 2020 Census to address disclosure risks from reconstruction and reidentification attacks \citep{USCB21b,ITA+22-en,AAC+22}. The large value $\varepsilon=19.61$ announced for the P.L.~94-171 redistricting data \citep{USCB21a} has raised questions about the adequacy of the privacy protection provided \citep{MS22,KKM+23}. Interpreting this value requires understanding both how it was calculated and how the underlying guarantee constrains an adversary's ability to distinguish neighboring databases.

DP uses a parameter $\varepsilon$ to bound the ratio between the probabilities of any output event under two neighboring databases, with smaller values giving stronger guarantees \citep{Dwo06}. These constraints can be interpreted through the relationship between the type I error probability (false positive rate) and power (true positive rate) of a hypothesis test that distinguishes the two output distributions \citep{WZ10,KOV17}. Extensions such as zCDP \citep{BS16} and Gaussian DP (GDP) \citep{DRS22} express guarantees using parameters other than $\varepsilon$ and facilitate the analysis of Gaussian mechanisms and composition.

These guarantees are commonly reported by converting them to approximate DP (ADP), which allows an additive relaxation $\delta$ in the DP inequality, and stating the resulting $\varepsilon$ for a fixed, small $\delta$ \citep{GKK+26a}. Summarizing a guarantee by an ADP pair with a small, fixed $\delta$ can, however, substantially loosen its power bound at a significance level of interest \citep[Section~2.3]{NDLH25}. In a report commissioned by the Census Bureau, JASON shows that the published pair $(\varepsilon,\delta)=(19.61,10^{-10})$ gives a power bound close to $1$ at a significance level of $1\%$ \citep{JAS22}.

The choice of guarantee format also affects comparisons of data utility between mechanisms. \citet[Section~4]{CKS22} point out that the relative utility of Laplace and Gaussian mechanisms can depend on the format used to match their guarantees. \citet[Section~V-A]{KOV17} also compare the noise variances sufficient for the Laplace and Gaussian mechanisms to satisfy a common ADP guarantee under composition. In official statistics, such comparisons have been made for NN-Wavelet, a method for releasing aggregate data \citep{IT25}, and for census tables \citep{ITK+25,ITK+26-en}. These studies convert the Gaussian mechanism's zCDP guarantee to ADP and match the resulting $\varepsilon$ to the Laplace mechanism's DP parameter $\varepsilon$. Interpreting the resulting utility comparisons requires understanding how the matched parameters constrain an adversary's ability to distinguish neighboring databases.

Building on the hypothesis-testing interpretation of privacy guarantees, we propose reporting their power bounds on a scale shared with DP's $\varepsilon$. At a significance level $0<\alpha<1$, which sets an upper limit on the false positive rate, we solve the standard DP testing bounds for $\varepsilon$ to obtain the smallest DP parameter giving the same power bound as the original guarantee. We call this value the \emph{$\alpha$-operationally equivalent $\varepsilon$}. Reporting it together with $\alpha$ allows the original guarantee's constraint to be compared with that of DP at the same significance level, without loosening the original constraint.

Our contributions are as follows.
\begin{enumerate}
\item We formulate a reporting method that represents general power-bound curves on the scale of DP's $\varepsilon$. We show that operationally equivalent $\varepsilon$, as a function of the significance level, represents the input power bound exactly on $(0,1)$, even when that bound is nonconcave. We also characterize its extrema over an interval of significance levels and interpret their finite values as the parameters of pure DP power bounds enclosing the input bound on that interval, providing a basis for concise interval reporting.
\item We quantify the slack introduced by reporting conversions while keeping each input power bound fixed. For Laplace and Gaussian mechanisms with equal noise variance, we show that single ADP pairs and GDP envelopes introduce different amounts of slack for the two mechanisms and can yield an ordering of the mechanisms opposite to that of their maximum-power curves at a specified significance level. Direct reporting with operationally equivalent $\varepsilon$ preserves the ordering of the maximum-power curves at each specified significance level.
\item We reassess the published guarantee for the 2020 U.S. Census P.L.~94-171 redistricting data. Using the budget documents and the public implementation, we reproduce the published $\varepsilon=19.61$ as the sum of rounded component ADP values and identify the conversion formula actually used. We also show that direct conversion of the composed zCDP budget to a single ADP pair at $\delta=10^{-10}$ substantially loosens the power bounds of the original guarantee, and quantify this slack on the DP scale using operationally equivalent $\varepsilon$. This reassessment clarifies the privacy protection that is difficult to discern from the large published value $\varepsilon=19.61$.
\end{enumerate}

\section{Hypothesis-Testing Interpretations of Privacy Guarantees and ADP Reporting}

Privacy guarantees constrain an adversary's ability to distinguish neighboring databases by bounding test power at specified significance levels. We review this interpretation for DP and its generalization $f$-DP, represent guarantees by power bounds, and derive the bounds for DP, ADP, zCDP, and GDP\@. We then examine the drawbacks of reporting these guarantees through ADP conversion and the alternatives that motivate our proposal.

\subsection{DP and ADP}\label{sec:dp-adp}

Let $\mathcal{D}$ be a set of databases, $\mathcal{Y}$ an output space, and $\sim$ a symmetric adjacency relation on $\mathcal{D}$. A mechanism is a randomized map $M:\mathcal{D}\to\mathcal{Y}$.

For $\varepsilon\ge0$, a mechanism $M$ satisfies $\varepsilon$-DP\footnote{DP in this sense is also called pure DP to distinguish it from its extensions.} if, for every $D_0\sim D_1$ and every measurable $S\subseteq\mathcal{Y}$,
\[
\Pr[M(D_0)\in S] \le e^\varepsilon\Pr[M(D_1)\in S]
\]
holds \citep{Dwo06}.

ADP extends DP by allowing an additive relaxation $\delta$ in this inequality. For $\varepsilon\ge0$ and $0\le\delta<1$, a mechanism satisfies $(\varepsilon,\delta)$-ADP if
\[
\Pr[M(D_0)\in S] \le e^\varepsilon\Pr[M(D_1)\in S]+\delta
\]
holds for every neighboring pair and measurable set $S$ \citep{DKM+06}. By definition, $(\varepsilon,0)$-ADP is equivalent to $\varepsilon$-DP\@. Throughout this paper, we distinguish DP from ADP and refer to guarantees with $\delta>0$ as ADP guarantees.

DP and ADP are closed under composition. Suppose that each mechanism satisfies its stated guarantee for every fixed sequence of preceding outputs. A guarantee for the composition is then obtained by adding the $\varepsilon$ parameters for DP, or by adding the $\varepsilon$ and $\delta$ parameters separately for ADP.

\subsection{Operational Interpretation through Hypothesis Testing}\label{sec:hypothesis-testing}

The operational meaning of DP can be expressed through a test that distinguishes neighboring databases \citep{WZ10}. For $D_0\sim D_1$, let $P_0$ and $P_1$ be the output distributions of $M(D_0)$ and $M(D_1)$, respectively. Given an output $Y$, an adversary tests the null hypothesis $H_0:D=D_0$ against the alternative $H_1:D=D_1$. A test function $\phi:\mathcal{Y}\to[0,1]$ rejects $H_0$ with probability $\phi(y)$ upon observing $y$. Its type I error probability, type II error probability, and power are
\[
\alpha_\phi=\mathbb{E}_{P_0}[\phi(Y)],\qquad
\beta_\phi=\mathbb{E}_{P_1}[1-\phi(Y)],\qquad
\pi_\phi=1-\beta_\phi.
\]
The type I and type II error probabilities are the false positive and false negative rates, respectively, and power is the true positive rate. We use the term \emph{significance level} $\alpha$ for an upper limit on the type I error probability and consider tests satisfying $\alpha_\phi\le\alpha$.

Requiring
\[
\pi_\phi\le e^\varepsilon\alpha_\phi,\qquad
\pi_\phi\le1-e^{-\varepsilon}(1-\alpha_\phi)
\]
for every neighboring database pair and every test is equivalent to $\varepsilon$-DP \citep[Theorem~2 and Remark~3]{KOV17}. Thus, DP bounds the maximum power at each significance level. When the test is clear from context, we omit the subscript $\phi$ on the type II error probability and power.

\citet{DRS22} generalize this formulation to $f$-DP\@. For output distributions $P$ and $Q$ under the null and alternative hypotheses, respectively, define the function $T(P,Q)\colon[0,1]\to[0,1]$ by
\[
T(P,Q)(\alpha)=\inf_{\phi:\,\alpha_\phi\le\alpha}\beta_\phi.
\]
The value $T(P,Q)(\alpha)$ is the smallest achievable type II error probability when testing $P$ against $Q$ with type I error probability at most $\alpha$. A function is called a \emph{trade-off function} if it can be expressed as $T(P,Q)$ for some pair of probability distributions on a common space. Every such function is continuous and convex on $[0,1]$.

An $f$-DP guarantee uses a trade-off function $f$ as a common lower bound on the trade-off functions $T(M(D_0),M(D_1))$ for all ordered neighboring database pairs. Specifically, a mechanism $M$ satisfies $f$-DP if
\[
T(M(D_0),M(D_1))(\alpha)\ge f(\alpha)\qquad(0\le\alpha\le1)
\]
for every ordered pair $(D_0,D_1)$ with $D_0\sim D_1$ \citep{DRS22}. Because the adjacency relation is symmetric, this condition covers both testing directions.

Note that $f$, being a trade-off function, must be convex, whereas the pointwise infimum of the trade-off functions over all ordered neighboring pairs need not be convex \citep[Section~2.2]{DRS22}. Appendix~\ref{app:nonconcave-power} gives a binary example. If this infimum is nonconvex, no trade-off function $f$ can coincide with this tight common lower bound at every significance level.

\subsection{Power Bounds as Representations of Guarantees}\label{sec:power-bound-representation}

The DP constraints in Section~\ref{sec:hypothesis-testing} bound power, the adversary's true positive rate, which measures the success of a distinguishing attack directly. We therefore express privacy constraints in terms of power rather than type II error probability. For a fixed pair of distributions $P,Q$, the relation $\pi=1-\beta$ gives the maximum power at significance level $\alpha$ as
\[
G^{P,Q}(\alpha)=1-T(P,Q)(\alpha).
\]
Taking the supremum over all neighboring database pairs gives the \emph{maximum-power curve} of $M$:
\[
G^M(\alpha)=\sup_{D_0\sim D_1}G^{M(D_0),M(D_1)}(\alpha).
\]
The supremum ranges over ordered neighboring pairs, including both testing directions.

We represent a privacy guarantee by a function $g:[0,1]\to[0,1]$ that bounds power at each significance level. A mechanism $M$ satisfies the guarantee represented by $g$ if
\[
G^M(\alpha)\le g(\alpha)
\]
for every $\alpha\in[0,1]$. For any neighboring pair, the power of a test whose type I error probability is at most $\alpha$ is then at most $g(\alpha)$.

The two constraints in Section~\ref{sec:hypothesis-testing} give the power bound for $\varepsilon$-DP:
\begin{equation}
g_\varepsilon^{\mathrm{DP}}(\alpha)
=\min\left\{e^\varepsilon\alpha,\ 1-e^{-\varepsilon}(1-\alpha)\right\}.
\label{eq:pure-dp-power}
\end{equation}
A mechanism $M$ satisfies $\varepsilon$-DP if and only if $G^M(\alpha)\le g_\varepsilon^{\mathrm{DP}}(\alpha)$ for every $0\le\alpha\le1$. Similarly, the power bound for $(\varepsilon,\delta)$-ADP is
\begin{equation}
g_{\varepsilon,\delta}^{\mathrm{ADP}}(\alpha)
=\min\left\{1,\ e^\varepsilon\alpha+\delta,\ 1-e^{-\varepsilon}(1-\alpha-\delta)\right\}.
\label{eq:adp-power}
\end{equation}

\paragraph{Relation to $f$-DP}
An $f$-DP guarantee is represented by $g^f(\alpha)=1-f(\alpha)$, but for a general power bound $g$, the function $1-g$ need not be a trade-off function. The function $1-f$ is necessarily concave since $f$ is convex, whereas a mechanism's maximum-power curve $G^M$ need not be concave (see the discussion at the end of Section~\ref{sec:hypothesis-testing}). This broader representation allows us to give a tight power bound by taking $g=G^M$ whenever the exact curve for $M$ is available.

\subsection{zCDP and GDP}\label{sec:gaussian-mechanism}

zCDP \citep{BS16} and GDP \citep{DRS22} are extensions of DP suited to Gaussian mechanisms and their compositions. The Gaussian mechanism adds independent Gaussian noise with standard deviation $\sigma>0$ to each component of a query output \citep{DKM+06}. Let $\Delta$ denote the query's $L_2$ sensitivity, the supremum of the distance between its outputs on neighboring databases, and assume $0<\Delta<\infty$. This mechanism does not satisfy DP for any finite $\varepsilon$: for two output distributions with different means, tail events can have arbitrarily large probability ratios even as their probabilities under both distributions approach zero. For any $0<\delta<1$, however, it satisfies $(\varepsilon,\delta)$-ADP for a suitable finite $\varepsilon$.

zCDP extends DP using R\'enyi divergence. Write $D_\lambda(P\Vert Q)$ for the R\'enyi divergence of order $\lambda>1$. For $\rho\ge0$, a mechanism satisfies $\rho$-zCDP\footnote{The general definition is $(\xi,\rho)$-zCDP; $\rho$-zCDP abbreviates $(0,\rho)$-zCDP\@. Gaussian mechanisms and the Census guarantee in Section~\ref{sec:census} are of this form, so we use only $\rho$.} if $D_\lambda(M(D_0)\Vert M(D_1))\le\rho\lambda$ for every $D_0\sim D_1$ and every $\lambda>1$.

A $\rho$-zCDP guarantee also bounds power at each significance level through the hypothesis-testing interpretation that \citet{BBG+20} developed for R\'enyi DP\@. A test maps the mechanism's output to rejection or non-rejection, and R\'enyi divergence does not increase under this post-processing. Let $\operatorname{Ber}(x)$ denote the Bernoulli distribution with success probability $x$. A test with type I error probability $\alpha$ and power $\pi$ therefore satisfies, for every $\lambda>1$,
\[
D_\lambda(\operatorname{Ber}(\pi)\Vert\operatorname{Ber}(\alpha))\le\rho\lambda,
\qquad
D_\lambda(\operatorname{Ber}(\alpha)\Vert\operatorname{Ber}(\pi))\le\rho\lambda.
\]
For each order and direction, the constraint bounds $\pi$ from above. We define $g_\rho^{\mathrm{zCDP}}(\alpha)$ as the infimum of these bounds over $\lambda>1$ and both directions. This is the tightest pointwise bound implied by $\rho$-zCDP alone, without additional information about the mechanism's construction \citep{RGC+26}.

GDP is a special case of $f$-DP\@. Let $\Phi$ be the standard normal cumulative distribution function, with $\Phi^{-1}(0)=-\infty$ and $\Phi^{-1}(1)=+\infty$. For $\mu\ge0$, define
\[
T_\mu(\alpha)=\Phi\!\left(\Phi^{-1}(1-\alpha)-\mu\right).
\]
Then $\mu$-GDP is $f$-DP with $f=T_\mu$, and its power bound is
\begin{equation}
g_\mu^{\mathrm{GDP}}(\alpha)=1-T_\mu(\alpha)
=\Phi\!\left(\Phi^{-1}(\alpha)+\mu\right).
\label{eq:gdp-power}
\end{equation}
The Gaussian mechanism satisfies both guarantees: $\rho$-zCDP with $\rho=\Delta^2/(2\sigma^2)$, and $\mu$-GDP with $\mu=\Delta/\sigma$. Its maximum-power curve is $g_\mu^{\mathrm{GDP}}$, so the GDP guarantee is exact for this mechanism. The two parameters are related by $\mu^2=2\rho$. This relation holds for the Gaussian mechanism; applying it to another mechanism requires a separate justification of its GDP guarantee or Gaussian approximation.

Under the composition conditions of Section~\ref{sec:dp-adp}, mechanisms satisfying $\rho_i$-zCDP compose to give $\rho$-zCDP with $\rho=\sum_i\rho_i$ \citep{BS16}, and mechanisms satisfying $\mu_i$-GDP compose to give $\mu$-GDP with $\mu=\sqrt{\sum_i\mu_i^2}$ \citep{DRS22}. The discrete Gaussian mechanism, which adds discrete Gaussian noise on the integer lattice to an integer-valued query, also admits a zCDP guarantee \citep{CKS22}.

\subsection{Reporting through ADP Conversion and Its Alternatives}\label{sec:adp-reporting}

zCDP and GDP express privacy guarantees in terms of $\rho$ and $\mu$, respectively. Their numerical values are not directly comparable with $\varepsilon$. It is customary to convert these guarantees to ADP with a ``negligibly'' small $\delta$ and report the resulting $\varepsilon$ as a measure of protection. The 2020 U.S. Census budget allocation document follows this practice, converting the zCDP accounting results to ADP at $\delta=10^{-10}$ \citep{USCB21f}.

Several conversions give an $\varepsilon$ for which $\rho$-zCDP implies $(\varepsilon,\delta)$-ADP, including two by \citet[Lemmas~3.5 and~3.6]{BS16} and one by \citet[Corollary~13]{CKS22}. For fixed $\delta\in(0,1)$, the last conversion, which is tighter than the other two, gives
\begin{equation}
\varepsilon(\rho;\delta)=\max\left\{0,\inf_{\lambda>1}\left[
\rho\lambda+\log(1-1/\lambda)-\frac{\log(\lambda\delta)}{\lambda-1}
\right]\right\}.
\label{eq:zcdp-adp-conversion}
\end{equation}
For $\mu$-GDP with $\mu>0$, \citet[Corollary~1]{DRS22} give the smallest $\delta$ for a given $\varepsilon\ge0$ as
\begin{equation}
\delta_\mu(\varepsilon)
=\Phi\!\left(\frac{\mu}{2}-\frac{\varepsilon}{\mu}\right)
-e^\varepsilon\Phi\!\left(-\frac{\mu}{2}-\frac{\varepsilon}{\mu}\right),
\label{eq:gdp-adp-profile}
\end{equation}
and the family of all such pairs is equivalent to the GDP guarantee.

Summarizing a guarantee by an $(\varepsilon,\delta)$ pair with such a negligibly small $\delta$, however, has the following drawbacks.
\begin{description}
\item[Loss of the original constraints] Substituting the converted pair into~\eqref{eq:adp-power} can give a power bound for distinguishing attacks that is substantially looser than the bound from the original guarantee. Indeed, the pair $(19.61,10^{-10})$ reported as the privacy guarantee for P.L.~94-171 gives power bounds close to $1$, which impose virtually no constraint on distinguishing power.
\item[Discretion in the choice of $\delta$] There is no absolute criterion for what counts as a negligible $\delta$, and several criteria have been used \citep{GKK+26a}. Even for the same original guarantee, the reported $\varepsilon$ depends on the chosen $\delta$, which leaves room for the reporting party to choose $\delta$ so as to obtain a smaller $\varepsilon$. Readers who focus only on $\varepsilon$ may mistake the effect of this choice for a difference in the underlying protection.
\item[Loss of information about the exclusion of catastrophic failure] A $\rho$-zCDP or $\mu$-GDP guarantee excludes \emph{catastrophic failure}, meaning the existence of a distinguishing test with zero type I error probability and positive power for some neighboring database pair. This follows from $g_\rho^{\mathrm{zCDP}}(0)=g_\mu^{\mathrm{GDP}}(0)=0$. A single ADP pair with $\delta>0$, however, gives $g_{\varepsilon,\delta}^{\mathrm{ADP}}(0)=\delta>0$, so this ADP guarantee alone does not establish the exclusion of catastrophic failure.
\end{description}

Several alternatives to such a summary avoid some of these drawbacks. Reporting the parameter of the original guarantee, $\rho$ or $\mu$, retains all the information in that guarantee, but the parameter is not directly comparable with $\varepsilon$, as noted above. \citet{GKK+26a} propose summarizing the original guarantee by a tight, non-asymptotic GDP guarantee and reporting $\mu$. This gives a single value with a testing interpretation, but $\mu$ is likewise not on the scale of $\varepsilon$, and the resulting power bound, hereafter referred to as the \emph{GDP envelope}, can be loose for non-Gaussian mechanisms (Section~\ref{sec:reporting-slack}).\footnote{When GDP provides a poor fit or no GDP guarantee is available, \citet{GKK+26a} recommend reporting the tightest available guarantee, such as a tabulated privacy profile or trade-off curve, in addition to $\varepsilon$ at fixed $\delta$.}

Reporting the full power-bound curve $g$ preserves the power bound at every significance level, as does reporting an equivalent trade-off function $f$ or full privacy profile when such a representation exists.\footnote{A full privacy profile $\varepsilon(\delta)$ gives, for each $\delta$, the smallest $\varepsilon$ for which $(\varepsilon,\delta)$-ADP follows from the guarantee. A symmetric trade-off function $f$ and its corresponding full privacy profile represent the same privacy guarantee \citep[Proposition~6]{DRS22}.} Such a report represents the guarantee as a function rather than a single value, which complicates interpretation and comparison, as NIST guidance explains \citep[Section~2.3]{NDLH25}.

Reporting the power bound at specified significance levels, as in the operational evaluations of \citet{KAA+22} and \citet{JAS22}, gives values with a direct testing interpretation; these values are probabilities, however, and are not directly comparable with the DP parameter $\varepsilon$ used in existing guidance and comparisons. The next section combines this last approach with the DP scale: for a specified significance level, we express the power bound as the DP parameter that gives the same bound.

\section{\texorpdfstring{$\alpha$-Operationally Equivalent $\varepsilon$}{Alpha-Operationally Equivalent Epsilon}}\label{sec:operational-epsilon}

This section defines the $\alpha$-operationally equivalent $\varepsilon$, denoted by $\varepsilon_\alpha$, and illustrates its meaning with a GDP example. We then establish its basic properties and explain what information is retained by the full curve and by interval summaries.

\begin{defi}[$\alpha$-operationally equivalent $\varepsilon$]\label{def:operational-epsilon}
Let $g:[0,1]\to[0,1]$ represent a guarantee and satisfy $\alpha\le g(\alpha)\le1$ for every $\alpha\in[0,1]$. For each $\alpha\in[0,1]$, define
\[
\varepsilon_\alpha(g)=\inf\left\{\varepsilon\ge0:
g_\varepsilon^{\mathrm{DP}}(\alpha)\ge g(\alpha)\right\}.
\]
If no finite $\varepsilon$ satisfies the condition, set $\varepsilon_\alpha(g)=\infty$. We abbreviate this value as $\varepsilon_\alpha$ when $g$ is clear from context.
\end{defi}

Definition~\ref{def:operational-epsilon} and~\eqref{eq:pure-dp-power} yield the closed-form expression
\begin{equation}
\begin{aligned}
\varepsilon_\alpha(g)&=
\begin{cases}
0, & \alpha=0,\ g(0)=0,\\
\infty, & \alpha=0,\ g(0)>0,\\
h_\alpha(g), & 0<\alpha<1,\ g(\alpha)<1,\\
\infty, & 0<\alpha<1,\ g(\alpha)=1,\\
0, & \alpha=1,
\end{cases}\\
h_\alpha(g)&=\max\left\{\log\frac{g(\alpha)}{\alpha},\
\log\frac{1-\alpha}{1-g(\alpha)}\right\}.
\end{aligned}
\label{eq:operational-epsilon}
\end{equation}
The two logarithmic terms in~\eqref{eq:operational-epsilon} result from solving the two DP power constraints in~\eqref{eq:pure-dp-power} for $\varepsilon$. Their maximum, $h_\alpha(g)$, ensures that both constraints hold.\footnote{The two logarithmic terms have the same form as those in the $\delta=0$ case of the lower-bound estimator used in privacy auditing \citep{NST+21,NHS+23}. Section~\ref{sec:related-work} discusses the difference in how the expression is used.}

A point $\alpha_*\in(0,1)$ satisfying $g(\alpha_*)=1-\alpha_*$ is called a \emph{switching point}. For $\alpha<g(\alpha)<1$, the first logarithmic term in~\eqref{eq:operational-epsilon} is larger when $g(\alpha)<1-\alpha$, and the second is larger when $g(\alpha)>1-\alpha$. The two terms are equal at a switching point.

The value $\varepsilon_0(g)$ indicates whether the bound at $\alpha=0$ rules out catastrophic failure. If $g(0)=0$, then $\varepsilon_0(g)=0$ and such failure is excluded. If $g(0)>0$, then $\varepsilon_0(g)=\infty$: the bound at this point permits positive power at zero type I error probability, but does not imply that a test with these properties exists. Every positive value of $g(0)$ maps to $\infty$, so $\varepsilon_0(g)$ retains only this distinction, not the magnitude of $g(0)$.

The equivalence in Definition~\ref{def:operational-epsilon} concerns only the power bound at the specified $\alpha$. A mechanism satisfying the guarantee represented by $g$ need not satisfy $\varepsilon_\alpha(g)$-DP\@. An $\varepsilon$-DP guarantee requires the DP bound with the same $\varepsilon$ to hold at every significance level.

\begin{figure}[tbp]
\centering
\includegraphics[width=\linewidth]{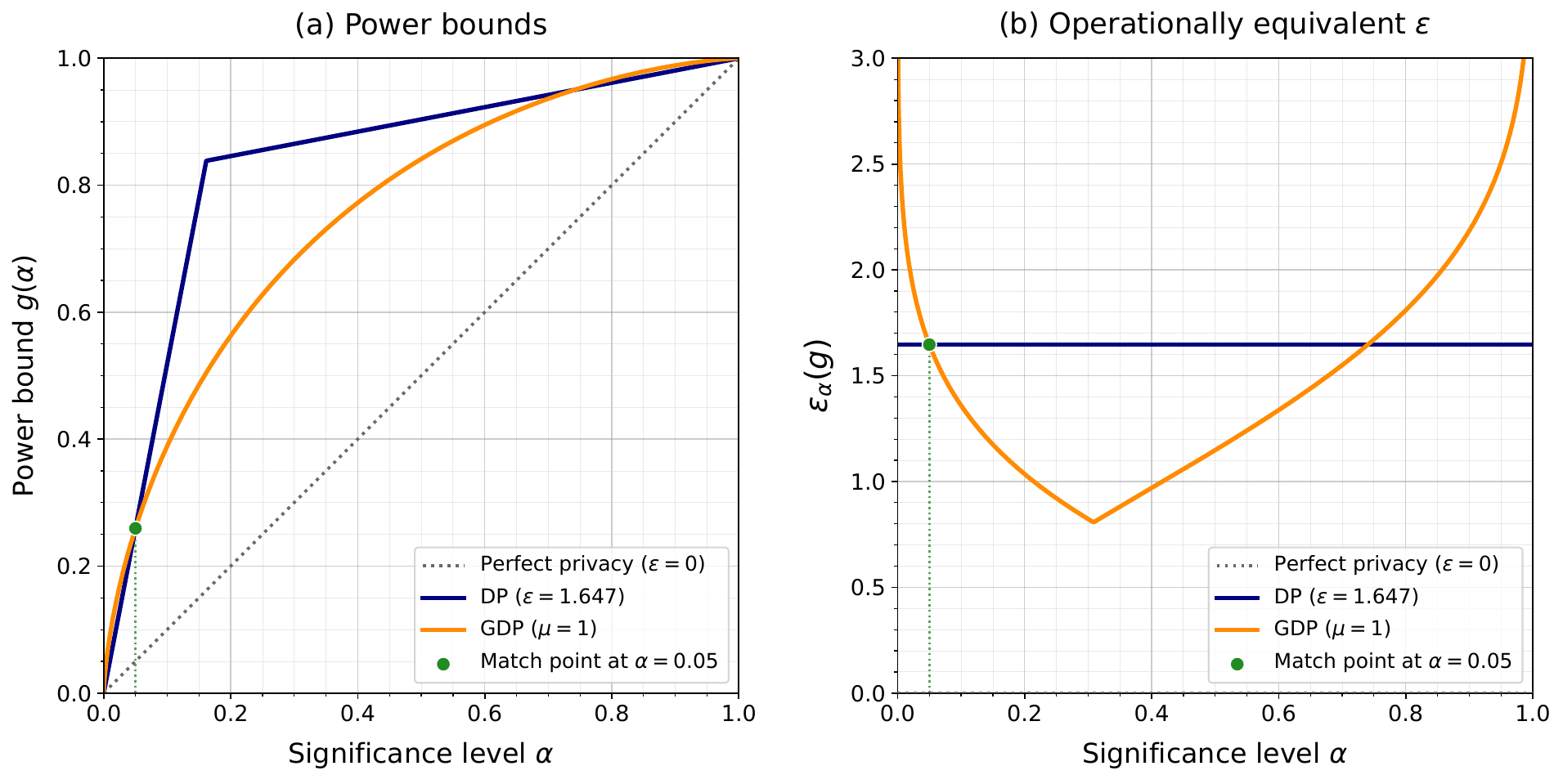}
\caption{Power bounds (a) and corresponding operationally equivalent $\varepsilon$ curves (b) for $1$-GDP and DP with $\varepsilon=\varepsilon_{0.05}(g_1^{\mathrm{GDP}})\approx1.647$. Green markers indicate the matching point at $\alpha=0.05$. Panel (b) shows the curves on $(0,1)$; the endpoint values $\varepsilon_0=\varepsilon_1=0$ are omitted.}
\label{fig:tradeoff-shapes}
\end{figure}

To illustrate the meaning of operational equivalence at a specified significance level, consider a $\mu$-GDP guarantee. For $0<\alpha<1$, substituting~\eqref{eq:gdp-power} into~\eqref{eq:operational-epsilon} gives
\[
\varepsilon_\alpha\!\left(g_\mu^{\mathrm{GDP}}\right)
=\max\left\{\log\frac{\Phi\!\left(\Phi^{-1}(\alpha)+\mu\right)}{\alpha},\
\log\frac{1-\alpha}{1-\Phi\!\left(\Phi^{-1}(\alpha)+\mu\right)}\right\}.
\]
With $\mu=1$ and $\alpha=0.05$, the power bound is approximately $0.2595$. The DP bound with $\varepsilon\approx1.647$ has the same value at this point, so $\varepsilon_{0.05}(g_1^{\mathrm{GDP}})\approx1.647$ (Figure~\ref{fig:tradeoff-shapes}(a)). Figure~\ref{fig:tradeoff-shapes}(b) shows the corresponding operationally equivalent $\varepsilon$ curves. The DP curve is constant on $(0,1)$, whereas the GDP curve varies with $\alpha$, illustrating that matching the bounds at a specified significance level does not imply that the full curves coincide.

The following proposition states the basic properties of $\varepsilon_\alpha(g)$ and characterizes its extrema over intervals of significance levels.

\begin{prop}[Basic properties]\label{prop:basic-properties}
The $\alpha$-operationally equivalent $\varepsilon$ has the following properties:
\begin{enumerate}[label=(\roman*)]
\item Consistency with DP guarantees: for every $\varepsilon\ge0$,
\[
\varepsilon_\alpha\!\left(g_\varepsilon^{\mathrm{DP}}\right)
=\begin{cases}
\varepsilon,&0<\alpha<1,\\
0,&\alpha\in\{0,1\}.
\end{cases}
\]
Moreover, for every mechanism $M$,
\[
\sup_{0<\alpha<1}\varepsilon_\alpha(G^M)
=\inf\{\varepsilon\ge0:M\text{ satisfies }\varepsilon\text{-DP}\},
\]
where $\inf\emptyset=\infty$.
\item Invertibility and equality of bounds: for each fixed $0<\alpha<1$, the map $g(\alpha)\mapsto\varepsilon_\alpha(g)$ is a bijection from $[\alpha,1]$ onto $[0,\infty]$. Its inverse is given by
\[
g(\alpha)=
\begin{cases}
g_{\varepsilon_\alpha(g)}^{\mathrm{DP}}(\alpha), & \varepsilon_\alpha(g)<\infty,\\
1, & \varepsilon_\alpha(g)=\infty.
\end{cases}
\]
The identity in the first case also holds at $\alpha\in\{0,1\}$ whenever $\varepsilon_\alpha(g)$ is finite.
\item Identification of $0$-DP: if $M$ satisfies the guarantee represented by $g$ and $g(\alpha_0)=\alpha_0$ for some $\alpha_0\in(0,1)$, then $M$ satisfies $0$-DP.
\item Order preservation: if $g_1(\alpha)\le g_2(\alpha)$, then $\varepsilon_\alpha(g_1)\le\varepsilon_\alpha(g_2)$.
\item Extrema over an interval: suppose $g=1-f$ for a trade-off function $f$, or $g=G^M$. For $0<a<b<1$, the maximum of $\varepsilon_\alpha(g)$ over $[a,b]$ is attained at an endpoint:
\[
\max_{\alpha\in[a,b]}\varepsilon_\alpha(g)
=\max\{\varepsilon_a(g),\varepsilon_b(g)\}.
\]
If there exists a switching point $\alpha_*\in[a,b]$ satisfying $g(\alpha_*)=1-\alpha_*$, then
\[
\min_{\alpha\in[a,b]}\varepsilon_\alpha(g)
=\varepsilon_{\alpha_*}(g)
=\log\frac{1-\alpha_*}{\alpha_*}.
\]
Otherwise, the minimum is attained at an endpoint.
\item Post-processing: for any post-processing $A$, $\varepsilon_\alpha(G^{A\circ M})\le\varepsilon_\alpha(G^M)$.
\end{enumerate}
\end{prop}

The proof is given in Appendix~\ref{app:properties-proofs}.

By Proposition~\ref{prop:basic-properties}(ii), the full curve $\alpha\mapsto\varepsilon_\alpha(g)$ determines $g(\alpha)$ at every significance level $0<\alpha<1$, without requiring concavity of $g$. In particular, the result applies to the maximum-power curve $G^M$ of any mechanism. If $g$ is right-continuous at $0$, the recovered interior values also determine $g(0)$; since $g(1)=1$, the reconstruction then extends to $[0,1]$. For $g=G^M$, Lemma~\ref{lem:zero-endpoint-recovery} gives an alternative formula for recovering $G^M(0)$ under this continuity assumption. Section~\ref{sec:reporting-comparison} compares how this method, $f$-DP, and full privacy profiles represent maximum-power curves.

The result on interval extrema in Proposition~\ref{prop:basic-properties}(v) also applies when the input is a nonconcave maximum-power curve. Lemma~\ref{lem:maximum-power-ratios} shows that $G^M$ is continuous on $(0,1)$ and that the ratios
\[
\frac{G^M(\alpha)}{\alpha},
\qquad
\frac{1-G^M(\alpha)}{1-\alpha}
\]
are both nonincreasing on this interval.\footnote{These ratio properties are necessary for a power bound to be the maximum-power curve of some mechanism. If a bound $g$ violates either property on $(0,1)$, then for every mechanism $M$ satisfying $G^M\le g$, there is some $\alpha\in(0,1)$ at which $G^M(\alpha)<g(\alpha)$.} Consequently, with $g=G^M$, the first logarithmic term in~\eqref{eq:operational-epsilon} is nonincreasing and the second is nondecreasing in $\alpha$. Together with continuity, these properties yield the endpoint and switching-point characterization in part~(v).

This characterization allows the range of $\varepsilon_\alpha(g)$ over a stated interval $[a,b]\subset(0,1)$ to be determined from the values at the endpoints and any switching point in the interval. Reporting this range provides a concise summary of the full curve $\alpha\mapsto\varepsilon_\alpha(g)$ without requiring a choice of individual evaluation points. Under the assumptions of Proposition~\ref{prop:basic-properties}(v), the extrema over $[a,b]$, where $0<a<b<1$, satisfy
\[
\begin{aligned}
\max_{\alpha\in[a,b]}\varepsilon_\alpha(g)
&=\inf\{\varepsilon\ge0: g_\varepsilon^{\mathrm{DP}}\ge g \text{ on }[a,b]\},\\
\min_{\alpha\in[a,b]}\varepsilon_\alpha(g)
&=\sup\{\varepsilon\ge0: g_\varepsilon^{\mathrm{DP}}\le g \text{ on }[a,b]\}.
\end{aligned}
\]
These identities follow from Definition~\ref{def:operational-epsilon}, Proposition~\ref{prop:basic-properties}(ii), and the strict monotonicity of $g_\varepsilon^{\mathrm{DP}}(\alpha)$ in $\varepsilon$ for $0<\alpha<1$. When the extrema are finite, the corresponding DP curves enclose $g$ on $[a,b]$, and each coincides with $g$ at some point in the interval.

The GDP example also illustrates how to compute an interval summary using Proposition~\ref{prop:basic-properties}(v). For $\mu>0$, the switching point is
\[
\alpha_*=\Phi(-\mu/2),\qquad
g_\mu^{\mathrm{GDP}}(\alpha_*)=1-\alpha_*.
\]
If the evaluation interval contains $\alpha_*$, the minimum of $\varepsilon_\alpha(g_\mu^{\mathrm{GDP}})$ over the interval is $\log((1-\alpha_*)/\alpha_*)$; otherwise, it is attained at an endpoint. The maximum is attained at an endpoint in either case. For $\mu=0$, $\varepsilon_\alpha=0$ at every $\alpha$.

An interval summary provides a concise report, but it does not in general determine the full curve $\alpha\mapsto\varepsilon_\alpha(g)$ and can discard constraints relevant to further composition. The enclosure on the stated interval also does not by itself imply a pure DP guarantee. A report intended to support further composition should therefore include the full operationally equivalent $\varepsilon$ curve or the original guarantee and its accounting parameters.

\section{Comparison with Other Reporting Formats}\label{sec:reporting-comparison}

We compare operationally equivalent $\varepsilon$ with the reporting alternatives discussed in Section~\ref{sec:adp-reporting}. Keeping each input power-bound curve fixed, we examine which constraints these formats preserve and how the slack introduced by reporting conversions affects comparisons between mechanisms.

\subsection{Expressiveness and Reporting Properties}\label{sec:reporting-properties}

Table~\ref{tab:reporting-formats} compares the reporting formats in terms of tightness, scalar reporting, and comparability with DP's $\varepsilon$.

\begin{table}[tp]
\caption{Expressiveness and reporting properties of privacy representations. Here, $G^M$ denotes the maximum-power curve of a mechanism $M$. For scalar and range reporting, auxiliary parameters are fixed and stated.}
\label{tab:reporting-formats}
\centering
\footnotesize
\setlength{\tabcolsep}{4pt}
\begin{tabular}{@{}p{0.24\linewidth}p{0.30\linewidth}p{0.16\linewidth}p{0.24\linewidth}@{}}
\toprule
{\raggedright Reporting format\par} & {\raggedright Tightness relative to $G^M$\par} & {\raggedright Scalar reporting\par} & {\raggedright Comparability with DP's $\varepsilon$\par} \\
\midrule
{\raggedright Single ADP pair with ``negligible'' $\delta$\par} & {\raggedright Not in general; can be substantially loose\par} & {\raggedright Yes, with $\delta$ fixed\par} & {\raggedright Approximately, if the effect of $\delta$ is negligible\par} \\
\addlinespace
{\raggedright $\mu$-GDP envelope\par} & {\raggedright Not in general (exact for Gaussian mechanisms)\par} & {\raggedright Yes\par} & {\raggedright Not directly\par} \\
\addlinespace
{\raggedright $f$-DP\par} & {\raggedright Exact throughout if $G^M$ is continuous and concave\par} & {\raggedright No\par} & {\raggedright Not directly\par} \\
\addlinespace
{\raggedright Full privacy profile $\varepsilon(\delta)$\par} & {\raggedright Same as $f$-DP\par} & {\raggedright No\par} & {\raggedright Approximately, if the effect of $\delta$ is negligible\par} \\
\addlinespace
{\raggedright Full power-bound curve $g$\par} & {\raggedright Exact throughout for any $G^M$\par} & {\raggedright No\par} & {\raggedright Not directly\par} \\
\addlinespace
{\raggedright Power bound $g(\alpha)$ at a specified $\alpha$\par} & {\raggedright Exact at the specified $\alpha$ for any $G^M$\par} & {\raggedright Yes, with $\alpha$ specified\par} & {\raggedright Not directly\par} \\
\midrule
{\raggedright Operationally equivalent $\varepsilon_\alpha(g)$ curve\par} & {\raggedright Exact on $(0,1)$ for any $G^M$ (exact throughout if $G^M(0)=0$ or $G^M$ is right-continuous at $0$)\par} & {\raggedright No\par} & {\raggedright Exactly at each significance level $0<\alpha<1$\par} \\
\addlinespace
{\raggedright Operationally equivalent $\varepsilon_\alpha(g)$ at a specified $\alpha$\par} & {\raggedright Exact at the specified $\alpha\in(0,1)$ for any $G^M$\par} & {\raggedright Yes, with $\alpha$ specified\par} & {\raggedright Exactly at the specified significance level\par} \\
\addlinespace
{\raggedright Interval summary of operationally equivalent $\varepsilon_\alpha(g)$\par} & {\raggedright Exact interval extrema on $[a,b]\subset(0,1)$ for any $G^M$\par} & {\raggedright Range, with the interval specified\par} & {\raggedright Exactly for finite extrema; these are the parameters of tight DP bounds enclosing $g$ on the interval\par} \\
\bottomrule
\end{tabular}
\end{table}

Single ADP pairs and GDP envelopes permit scalar reporting, but can loosen the input power bound. Comparing the ADP parameter $\varepsilon$ with DP's $\varepsilon$ relies on treating the effect of $\delta$ as negligible, whereas the GDP parameter $\mu$ is not directly comparable with $\varepsilon$.

Reporting a full curve can preserve power constraints lost in a single ADP pair or GDP envelope, but the absence of a scalar summary complicates comparison and communication (Section~\ref{sec:adp-reporting}). A full privacy profile, expressed as $\delta(\varepsilon)$ or its generalized inverse $\varepsilon(\delta)$, has the same expressive power as $f$-DP \citep[Propositions~2 and~6]{DRS22}. As discussed in Section~\ref{sec:power-bound-representation}, a general power-bound curve $g$ can represent $G^M$ tightly even when $f$-DP cannot. Appendix~\ref{app:nonconcave-power} gives an example in which both the tightest $f$-DP guarantee and the full privacy profile yield power bounds strictly larger than $G^M$ at some significance levels.

Trade-off functions and power-bound curves express constraints in terms of testing error probabilities or power, so their values are not directly comparable with DP's $\varepsilon$. In a full privacy profile, each $\varepsilon(\delta)$ is an ADP parameter; comparing it with DP's $\varepsilon$ still relies on treating the effect of $\delta$ as negligible.

The operationally equivalent $\varepsilon$ curve represents $g$ exactly on $(0,1)$ (Section~\ref{sec:operational-epsilon}). At a specified significance level $0<\alpha<1$, a single value $\varepsilon_\alpha(g)$ conveys the same information as $g(\alpha)$ (Proposition~\ref{prop:basic-properties}(ii)). When finite, this value is the DP parameter that yields the same power bound, allowing direct comparison with DP's $\varepsilon$ at that significance level. An interval summary gives the range of these DP parameters over a stated interval. Under the assumptions of Proposition~\ref{prop:basic-properties}(v), its finite extrema are the parameters of the DP curves that enclose $g$ most tightly on the interval. The summary does not in general determine the full curve.

\subsection{Slack and Ordering for Laplace and Gaussian Mechanisms}\label{sec:reporting-slack}

The following example illustrates how the differences in tightness discussed in Section~\ref{sec:reporting-properties} affect comparisons between mechanisms. We consider Laplace and Gaussian mechanisms with equal noise variance and compare their exact maximum-power curves with the power bounds conveyed by single ADP pairs and GDP envelopes. Both mechanisms add noise to the same scalar query of sensitivity $1$. The Laplace mechanism uses scale $b=1$, and the Gaussian mechanism uses standard deviation $\sigma=\sqrt{2}$, so both noise variances are $2$ and their expected squared errors coincide. We assume that the query values differ by $1$ for at least one pair of neighboring databases. The mechanisms' exact maximum-power curves, denoted by $G^{\mathrm{Lap}}$ and $G^{\mathrm{Gau}}$, are derived in Appendix~\ref{app:laplace-gaussian-computation}. In particular, $G^{\mathrm{Gau}}=g_{1/\sqrt{2}}^{\mathrm{GDP}}$.

\begin{figure}[tbp]
\centering
\includegraphics[width=\linewidth]{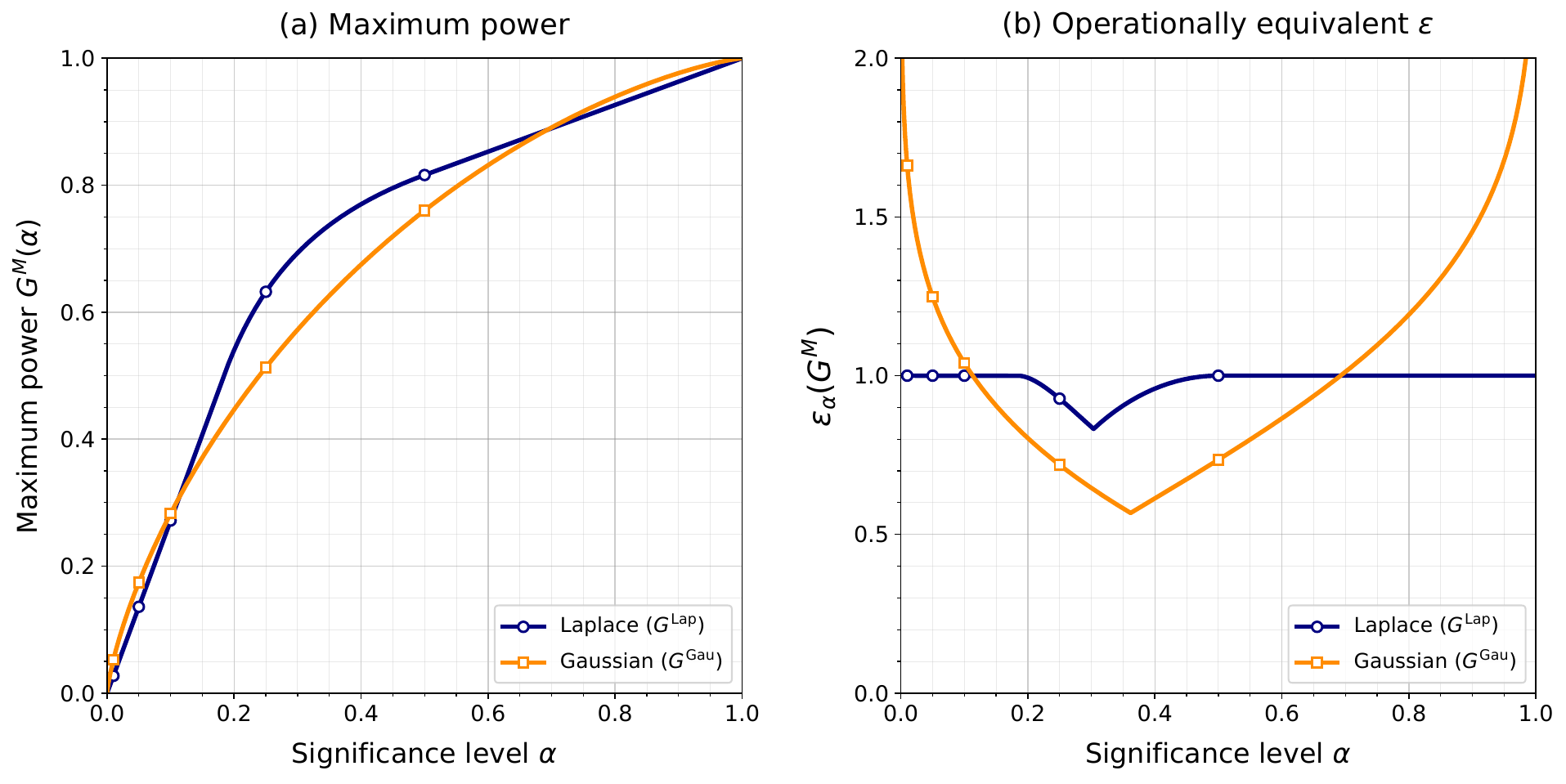}
\caption{Maximum-power curves (a) and corresponding operationally equivalent $\varepsilon$ curves (b) for Laplace and Gaussian mechanisms with sensitivity $1$ and noise variance $2$. The Laplace scale is $1$, and the Gaussian standard deviation is $\sqrt{2}$. Circles and squares mark the Laplace and Gaussian curves, respectively, at the five significance levels in Table~\ref{tab:laplace-gaussian-comparison}. Panel (b) shows the curves on $(0,1)$; the endpoint values $\varepsilon_0=\varepsilon_1=0$ are omitted.}
\label{fig:laplace-gaussian-power}
\end{figure}

The maximum-power curves cross, so their ordering depends on the significance level (Figure~\ref{fig:laplace-gaussian-power}(a)). The corresponding operationally equivalent $\varepsilon$ curves preserve this ordering at each significance level on the DP scale (Figure~\ref{fig:laplace-gaussian-power}(b)). Table~\ref{tab:laplace-gaussian-comparison} reports maximum power and its operationally equivalent $\varepsilon$ at five evaluation points. The Laplace mechanism has lower maximum power at $\alpha=0.01$, $0.05$, and $0.10$, whereas the Gaussian mechanism has lower maximum power at $\alpha=0.25$ and $0.50$.

\begin{table}[t]
\caption{Maximum power and its operationally equivalent $\varepsilon$ for mechanisms with equal noise variance $2$. The sensitivity is $1$, the Laplace scale is $1$, and the Gaussian standard deviation is $\sqrt{2}$. For each significance level, the smaller value in each pair of columns is shown in bold.}
\label{tab:laplace-gaussian-comparison}
\centering
\small
\begin{tabular}{@{}rrrrr@{}}
\toprule
& \multicolumn{2}{c}{Maximum power} & \multicolumn{2}{c}{Operationally equivalent $\varepsilon$} \\
\cmidrule(lr){2-3}\cmidrule(lr){4-5}
$\alpha$ & $G^{\mathrm{Lap}}(\alpha)$ & $G^{\mathrm{Gau}}(\alpha)$ & $\varepsilon_\alpha(G^{\mathrm{Lap}})$ & $\varepsilon_\alpha(G^{\mathrm{Gau}})$ \\
\midrule
0.01 & \textbf{0.027183} & 0.052698 & \textbf{1.0000} & 1.6620 \\
0.05 & \textbf{0.135914} & 0.174187 & \textbf{1.0000} & 1.2481 \\
0.10 & \textbf{0.271828} & 0.282833 & \textbf{1.0000} & 1.0397 \\
0.25 & 0.632121 & \textbf{0.513010} & 0.9276 & \textbf{0.7188} \\
0.50 & 0.816060 & \textbf{0.760250} & 1.0000 & \textbf{0.7350} \\
\bottomrule
\end{tabular}
\end{table}

Table~\ref{tab:format-ordering} shows how the same mechanisms are reported using single ADP pairs and GDP envelopes. For ADP reporting, we retain the Laplace mechanism's $1$-DP guarantee as $(1,0)$ and convert the Gaussian guarantee using~\eqref{eq:gdp-adp-profile} at the ``negligible'' value $\delta=10^{-10}$. For GDP reporting, we choose, for each mechanism, the smallest $\mu$ for which the GDP power curve bounds that mechanism's maximum-power curve throughout $[0,1]$. Appendix~\ref{sec:adp-point-selection} gives the ADP calculations, and Appendix~\ref{app:gdp-envelope} derives the minimum GDP envelopes.

\begin{table}[t]
\caption{ADP pairs and minimum GDP envelope parameters for the mechanisms in Table~\ref{tab:laplace-gaussian-comparison}. Bold entries indicate the stronger reported bound within each format.}
\label{tab:format-ordering}
\centering
\small
\begin{tabular}{@{}lrr@{}}
\toprule
Reporting format & Laplace & Gaussian \\
\midrule
Single ADP pair $(\varepsilon,\delta)$ & $\boldsymbol{(1,0)}$ & $(4.4922,10^{-10})$ \\
Minimum GDP envelope parameter $\mu$ & $1.0301$ & $\mathbf{0.7071}$ \\
\bottomrule
\end{tabular}
\end{table}

The two reporting formats give opposite orderings, and neither reproduces the maximum-power comparisons in Table~\ref{tab:laplace-gaussian-comparison}. The ADP reports give a smaller power bound for the Laplace mechanism at all five evaluation points, although the Gaussian mechanism has lower maximum power at the two largest significance levels. This reversal persists for all five small values of $\delta$ examined in Appendix~\ref{sec:adp-point-selection}. By contrast, the GDP envelopes give a smaller power bound for the Gaussian mechanism at all five points, although the Laplace mechanism has lower maximum power at the three smallest significance levels. ADP reporting substantially loosens the Gaussian bound while retaining the Laplace $1$-DP bound, whereas the GDP envelope is exact for the Gaussian mechanism but loose for the Laplace mechanism.

The slack at each evaluation point can be removed by choosing, for each mechanism, an ADP pair whose power bound is tangent to the corresponding maximum-power curve at that point (Appendix~\ref{sec:adp-point-selection}). For the Gaussian mechanism at $\alpha=0.50$, the pair $(\varepsilon,\delta)\approx(0.25,0.1922)$ reproduces the maximum power of approximately $0.7602$. Here, however, $\delta$ can no longer be treated as negligible. Both $\varepsilon$ and $\delta$ contribute to this power bound, so the pair's $\varepsilon$ component $0.25$ is not directly comparable with DP's $\varepsilon$. The DP parameter giving the same bound at $\alpha=0.50$ is $\varepsilon_{0.50}(G^{\mathrm{Gau}})\approx0.7350$.

\section{Reassessing Privacy Guarantees for the 2020 U.S. Census}\label{sec:census}

We use operationally equivalent $\varepsilon$ to reassess how the published guarantees for the P.L.~94-171 redistricting data constrain distinguishing attacks. We first verify how the published value $\varepsilon=19.61$ was calculated, then derive power bounds directly from the original zCDP guarantee and express them on a scale shared with DP\@. Comparison with the bounds obtained via a single ADP pair reveals protection that is difficult to discern from the large published value. We also provide a supplementary evaluation that approximates the discrete Gaussian noise-addition stage by continuous Gaussian mechanisms.

\subsection{Scope and Published Privacy Budgets}\label{sec:census-scope}

Our analysis concerns the published guarantees for the P.L.~94-171 data on persons and housing units. The TopDown Algorithm (TDA), the core algorithm of the Disclosure Avoidance System (DAS) used to produce these data, adds discrete Gaussian noise to counts and then post-processes them to enforce constraints such as consistency across geographic levels \citep{AAC+22}.

The scope of the guarantee depends on the adjacency relation and the treatment of invariants. TDA uses an adjacency relation in which neighboring databases differ by the replacement of one record. State population totals, housing-unit counts in each block, and the numbers and types of occupied group quarters are released exactly as invariants \citep[Sections~2.1 and~5.2]{AAC+22}. We evaluate the published guarantee for the noise-addition stage. For the final release, we consider tests that distinguish neighboring databases with identical invariants. For such databases, post-processing is the same function of the noisy counts, so the power bounds also apply to the released data.

The published budgets are $\rho_{\mathrm p}\approx2.56$ for persons, $\rho_{\mathrm h}\approx0.07$ for housing units, and $\rho\approx2.63$ for their composition. The budget allocation document reports corresponding ADP values of $\varepsilon=17.14$, $2.47$, and $17.44$, respectively, at $\delta=10^{-10}$ \citep{USCB21f}. The press release, however, reports $\varepsilon=19.61$ as the overall budget \citep{USCB21a}.

The computations below use the exact fractions recorded in the separate Certificates of Disclosure Avoidance for persons and housing units \citep{USCB21e}. Their composed budget is
\[
\rho=\rho_{\mathrm p}+\rho_{\mathrm h}=2.631169245\ldots.
\]
Appendix~\ref{app:tda-parameters} lists these fractions and shows how they are combined.

\subsection{Verifying the Published ADP Conversions}\label{sec:census-adp-conversions}

The conversion in~\eqref{eq:zcdp-adp-conversion} \citep[Corollary~13]{CKS22} reproduces all $\varepsilon$ values in the budget allocation document to the displayed precision, including the component values and the value after composition. \citet[Section~2.1]{AAC+22} state that the Bun--Steinke conversion was used. The public implementation \citep{USCB21d}, however, uses the Canonne et al.\ formula. Specifically, the budget calculation calls the conversion function with \texttt{zcdp=True}, and \path{BaseCurve.get_zcdp_delta()} evaluates that formula.\footnote{Public repository commit \texttt{289ee463936a}: \path{source/programs/engine/budget.py:129-155}, and \path{curve.py:170-199} (implementation of the conversion) and \path{curve.py:212-217} (selection of the conversion function) in the same directory.}

Table~\ref{tab:zcdp-adp-conversions} compares the two Bun--Steinke conversions \citep[Lemmas~3.5 and~3.6]{BS16} with~\eqref{eq:zcdp-adp-conversion}, using the budgets recorded in the certificates. In the computed rows, ``Overall'' is obtained by adding the $\rho$ values for persons and housing units before conversion.

\begin{table}[t]
\caption{ADP conversions for P.L.~94-171 with $\delta=10^{-10}$. Computed rows use the $\rho$ values recorded in the certificates; the ``Overall'' values are obtained by summing the component budgets before conversion. Published values retain their original precision.}
\label{tab:zcdp-adp-conversions}
\centering
\small
\begin{tabular}{@{}lrrr@{}}
\toprule
& \multicolumn{3}{c}{$\varepsilon$} \\
\cmidrule(l){2-4}
Source or conversion & Persons & Housing units & Overall \\
\midrule
Press release \citep{USCB21a} & 17.14 & 2.47 & 19.61 \\
Budget allocation \citep{USCB21f} & 17.14 & 2.47 & 17.44 \\
\citet[Lemma~3.5]{BS16} & 17.9002 & 2.7022 & 18.1984 \\
\citet[Lemma~3.6]{BS16} & 17.4361 & 2.5330 & 17.7313 \\
\citet[Corollary~13]{CKS22} & 17.1436 & 2.4747 & 17.4351 \\
\bottomrule
\end{tabular}
\end{table}

The value $19.61$ in the press release equals the sum of the displayed component values:
\[
17.14+2.47=19.61.
\]
Adding the unrounded converted values gives $19.6182\ldots$, which rounds to $19.62$. Thus, $19.61$ is reproduced by summing the rounded component values, whereas $17.44$ is obtained by converting the composed budget directly.

Basic composition of the component ADP guarantees adds $\delta$ as well as $\varepsilon$. Composing two guarantees each with $\delta=10^{-10}$ yields $\delta=2\times10^{-10}$, which differs from the fixed $\delta=10^{-10}$ used to convert the composed budget directly.

\subsection{Reporting from the zCDP Guarantee}\label{sec:census-zcdp}

The published composed budget $\rho=2.631169245\ldots$ yields operationally equivalent $\varepsilon$ values of approximately $4.25$, $2.94$, and $3.18$ at $\alpha=0.01$, $0.05$, and $0.10$, respectively (Table~\ref{tab:power-comparison}). They are obtained from $g_\rho^{\mathrm{zCDP}}$, the bound derived directly from zCDP in Section~\ref{sec:gaussian-mechanism}. We use the same significance levels as \citet[Section~6.3]{KAA+22} to enable comparison with their operational evaluation. At all three levels, the zCDP power bounds in Table~\ref{tab:power-comparison} agree to two decimal places with those in their Table~2.

\begin{table}[t]
\caption{Power bounds from the published zCDP guarantee and power values from the supplementary Gaussian approximation. Here $\rho=2.631169245\ldots$, and the GDP parameter used for the approximation is $\mu=\sqrt{2\rho}$. Displayed values are rounded.}
\label{tab:power-comparison}
\centering
\small
\begin{tabular}{@{}rrrrr@{}}
\toprule
& \multicolumn{2}{c}{zCDP guarantee} & \multicolumn{2}{c}{Gaussian approximation} \\
\cmidrule(lr){2-3}\cmidrule(lr){4-5}
$\alpha$ & Power bound & $\varepsilon_\alpha$ & Power & $\varepsilon_\alpha$ \\
\midrule
0.01 & 0.6984 & 4.25 & 0.4871 & 3.89 \\
0.05 & 0.9468 & 2.94 & 0.7419 & 2.70 \\
0.10 & 0.9624 & 3.18 & 0.8443 & 2.13 \\
\bottomrule
\end{tabular}
\end{table}

The value $\varepsilon_{0.01}\approx4.25$ expresses the power bound of approximately $0.6984$ on the DP scale: DP with $\varepsilon\approx4.25$ gives the same bound at a significance level of $1\%$. This value gives a more precise account of the constraint on distinguishing attacks than the bounds near $1$ obtained through ADP conversion.

The direct zCDP bound of approximately $0.6984$ at $\alpha=0.01$ is substantially tighter than JASON's reported bound of $0.997$. JASON calls the power bound at a significance level of $1\%$ the ``Effective TPR'' and reports $0.999999997$ for the published pair $(\varepsilon,\delta)=(19.61,10^{-10})$. It also searches for the smallest Effective TPR over ADP pairs converted from $\rho=2.63$, allowing $\delta$ to vary, and reports $0.997$ \citep[Table~5-7]{JAS22}. These conversions introduce slack, as the report itself notes. The R\'enyi divergence constraints for binary tests described in Section~\ref{sec:gaussian-mechanism} give the tightest pointwise power bound implied by zCDP alone. Applying these constraints directly avoids that conversion slack.

\begin{table}[t]
\caption{Power bounds implied by the ADP pair $(17.435109828\ldots,\,10^{-10})$ obtained from the published composed budget, compared with the direct zCDP bound $g$ of Table~\ref{tab:power-comparison}. Because the ADP bound $\widetilde g$ is close to $1$, we display $1-\widetilde g(\alpha)$.}
\label{tab:same-input-zcdp}
\centering
\small
\begin{tabular}{@{}rrrr@{}}
\toprule
$\alpha$ & $1-\widetilde g(\alpha)$ & $\varepsilon_\alpha(\widetilde g)$ & $\varepsilon_\alpha(\widetilde g)-\varepsilon_\alpha(g)$ \\
\midrule
0.01 & $2.65255\times10^{-8}$ & 17.4351 & 13.1889 \\
0.05 & $2.54537\times10^{-8}$ & 17.4351 & 14.4941 \\
0.10 & $2.41141\times10^{-8}$ & 17.4351 & 14.2600 \\
\bottomrule
\end{tabular}
\end{table}

The loss of information persists even if $19.61$ is replaced by $17.44$. Converting the composed budget with~\eqref{eq:zcdp-adp-conversion} at $\delta=10^{-10}$ gives $\varepsilon=17.435109828\ldots$, and substituting this pair into~\eqref{eq:adp-power} gives the bound $\widetilde g$. With the original zCDP guarantee held fixed, Table~\ref{tab:same-input-zcdp} compares the converted bound $\widetilde g$ with the direct bound $g=g_\rho^{\mathrm{zCDP}}$ at common evaluation points, using the same approach as in Appendix~\ref{sec:adp-point-selection}. The ADP bound is within $2.7\times10^{-8}$ of $1$ at every point, and its operationally equivalent $\varepsilon$ is approximately $17.44$, exceeding the direct values by $13.19$ to $14.49$. Reconstructing the published value explains how it was calculated; deriving bounds directly from zCDP reveals the constraints on distinguishing attacks that the original zCDP guarantee actually imposes.

\subsection{Supplementary Gaussian Approximation}\label{sec:census-gaussian}

To complement the power bounds derived from the published zCDP guarantee, we estimate distinguishing power by approximating the discrete Gaussian mechanisms with their continuous Gaussian counterparts. The GDP guarantee for a Gaussian mechanism and the GDP composition rule \citep{DRS22} give the parameter
\[
\mu=\sqrt{2\rho}=2.293978746\ldots.
\]
The values under ``Gaussian approximation'' in Table~\ref{tab:power-comparison} are computed from the power curve $g_\mu^{\mathrm{GDP}}$ with this parameter.

The Monte Carlo evaluation by \citet[Section~6.3]{KAA+22}, using the production budget allocation, suggests that this approximation closely reflects distinguishing power at the evaluated significance levels. They generated one million samples under each of the null and alternative hypotheses and estimated the power of likelihood-ratio tests on counts perturbed by discrete Gaussian noise. At all three significance levels in Table~\ref{tab:power-comparison}, their estimates and the values from the continuous Gaussian approximation agree when rounded to two decimal places.

The Gaussian approximation gives operationally equivalent $\varepsilon$ values of $3.89$, $2.70$, and $2.13$ at the three significance levels, respectively. Both evaluations in Table~\ref{tab:power-comparison} use the same mapping to $\varepsilon_\alpha$. Their numerical differences arise from how power is evaluated: one uses only the zCDP guarantee, whereas the other uses a Gaussian approximation of the noise-addition stage.

\begin{figure}[tp]
\centering
\includegraphics[width=\textwidth]{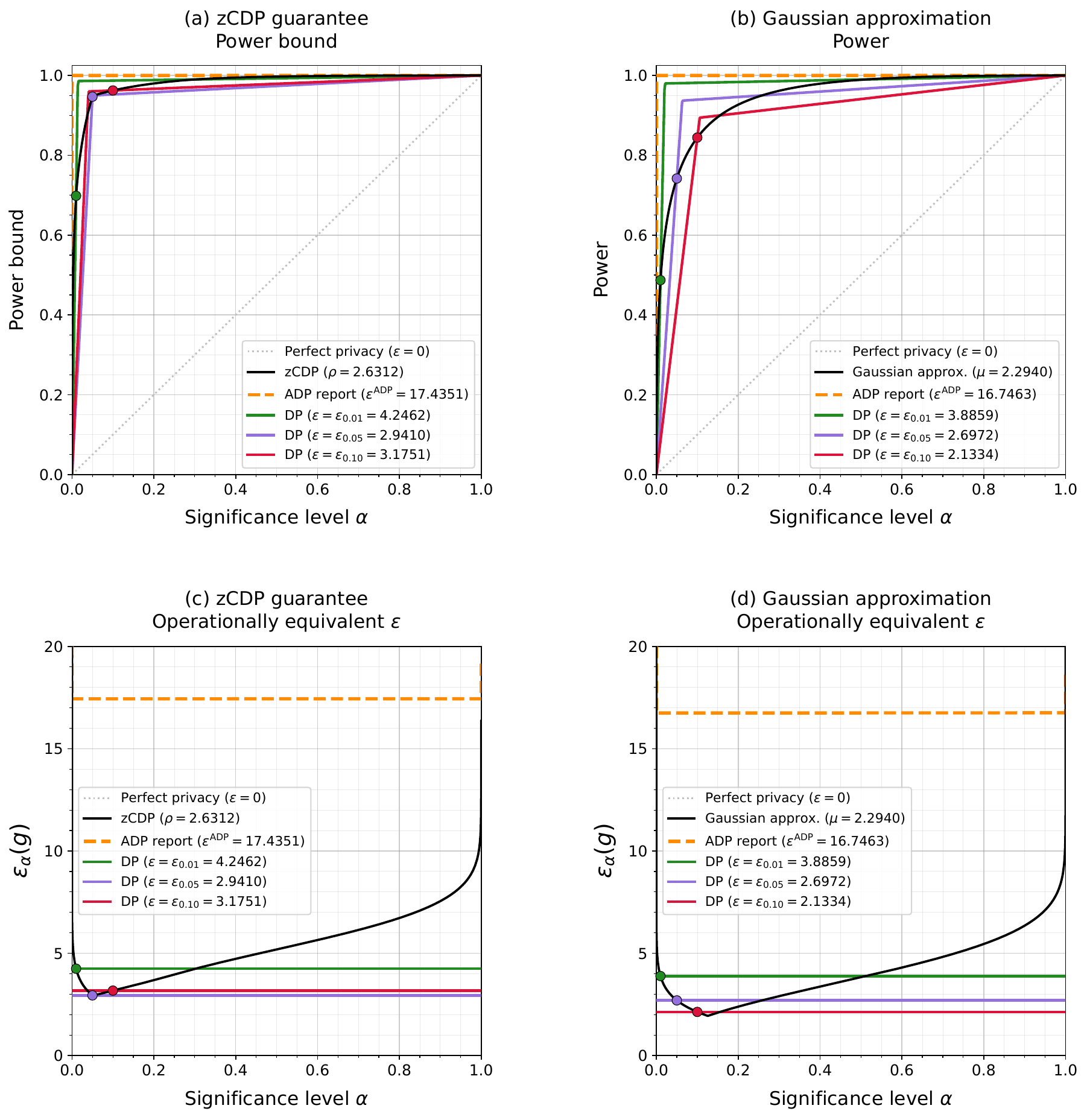}
\caption{Power and operationally equivalent $\varepsilon$ curves for the P.L.~94-171 reassessment. The black curves in (a) and (b) show the direct zCDP power bound for $\rho=2.631169245\ldots$ and the power curve under the supplementary Gaussian approximation with $\mu=\sqrt{2\rho}$, respectively. Panels (c) and (d) show the corresponding operationally equivalent $\varepsilon$ curves for $0<\alpha<1$. Orange dashed curves correspond to ADP pairs obtained at $\delta=10^{-10}$ via~\eqref{eq:zcdp-adp-conversion} (left column) and~\eqref{eq:gdp-adp-profile} (right column). Colored solid curves represent pure DP bounds, each matched to the corresponding black curve at one of the significance levels $\alpha=0.01$, $0.05$, and $0.10$. Circles mark the matching points.}
\label{fig:power-comparison}
\end{figure}

Figure~\ref{fig:power-comparison} presents the direct zCDP evaluation and the supplementary Gaussian approximation on both the power and operationally equivalent $\varepsilon$ scales, together with the corresponding ADP reports. At each marked significance level, the matching DP bound reproduces the power value from the corresponding evaluation, whereas the ADP bound is substantially looser.

\section{Discussion}

We first discuss what the reassessment in Section~\ref{sec:census} implies for interpreting Census privacy protection. We then consider how to summarize values that depend on the evaluation point and how conversion slack affects utility comparisons between mechanisms. Finally, we discuss the limitations of the evaluation and considerations for its use.

\subsection{Implications of the Census Reassessment}\label{sec:census-implications}

This reassessment shows that the original zCDP guarantee imposes substantially stronger constraints on distinguishing attacks than those conveyed by the reported ADP value $\varepsilon=19.61$. At the three significance levels evaluated in Section~\ref{sec:census-zcdp}, $\varepsilon_\alpha$ ranges from approximately $2.94$ to $4.25$, and the power bound at each level coincides with that of DP with the corresponding $\varepsilon$. This correspondence provides a clear quantitative basis for reconsidering the impression of weak privacy protection conveyed by the reported ADP value.

This change in interpretation follows from the same published zCDP budget. Conversion to a single ADP pair substantially loosens the original guarantee's constraints on distinguishability, presenting a picture of much weaker protection. The power bound close to $1$ obtained from the published ADP pair indicates that this pair conveys only weak constraints; it does not imply that the original guarantee imposes equally weak constraints. The effect of reporting conversion on the apparent protection must therefore be taken into account. We conclude that the magnitude of the published $\varepsilon=19.61$ should not be used as the main basis for judging the adequacy of Census privacy protection.

Reporting operationally equivalent $\varepsilon$ connects this reassessment to discussions of protection levels in terms of DP's $\varepsilon$. By matching each power bound from the original zCDP guarantee to that of DP at the same significance level, we can compare guarantees expressed in different formats in common operational terms. The adequacy of protection can then be assessed by comparing the constraints under the attack condition specified by each significance level with the degree of distinguishability considered acceptable for the protected entities and intended uses of the data. This reassessment thus provides a basis for discussing Census privacy protection in terms of the constraints imposed by the original guarantee.

\subsection{Dependence on Evaluation Points and Concise Reporting}\label{sec:choice-of-alpha}

For the same published zCDP guarantee evaluated in Section~\ref{sec:census}, operationally equivalent $\varepsilon$ varies with the evaluation point and is nonmonotone across the three points considered. Reporting a single value as representative of the entire guarantee can therefore obscure how the constraints on distinguishability vary with the significance level.

This dependence on $\alpha$ differs from the discretion in choosing $\delta$ discussed in Section~\ref{sec:adp-reporting}. The convention of choosing a negligibly small $\delta$ in ADP reporting has no clear meaning in terms of distinguishing attacks, and enlarging $\delta$ lowers the reported $\varepsilon$ while increasing the apparent risk of catastrophic failure, even though the original guarantee excludes such failure. The significance level $\alpha$, by contrast, specifies the adversary's maximum allowed false positive rate: it can be chosen to reflect the acceptable rate for the attack of interest. The value $\varepsilon_\alpha$ reproduces the input power bound at that level without adding conversion slack.

Reporting the range of $\varepsilon_\alpha$ over an interval of interest provides a concise summary of this dependence. Once the input bound and the evaluation interval are fixed, the minimum and maximum values are uniquely determined, so the reported range does not depend on a choice of individual evaluation points within the interval. The interval itself remains a reporting choice and should be stated together with the range to make clear which attack conditions the summary covers.

To communicate the exclusion of catastrophic failure, the report may also state $\varepsilon_0=0$, that is, $g(0)=0$, alongside evaluations at positive significance levels (Section~\ref{sec:operational-epsilon}).

Table~\ref{tab:report-example} illustrates this form of reporting for the Census reassessment. By Proposition~\ref{prop:basic-properties}(v), the ranges of $\varepsilon_\alpha$ for the zCDP bound and the Gaussian approximation are determined by the values at the endpoints and at any switching point within the evaluation interval.\footnote{Proposition~\ref{prop:basic-properties}(v) applies to the zCDP bound because $1-g_\rho^{\mathrm{zCDP}}$ is a trade-off function \citep[Corollary~4.2 and Remark~4.5]{RGC+26}.} For the zCDP guarantee, the minimum occurs at the switching point $\alpha_*\approx0.051$, and the maximum at $\alpha=0.01$. For the Gaussian approximation, the switching point $\alpha_*=\Phi(-\mu/2)\approx0.126$ exceeds the upper endpoint $0.10$ of the evaluation interval, so $\varepsilon_\alpha$ decreases throughout the interval.

\begin{table}[t]
\caption{Interval summary of the P.L.~94-171 reassessment, in the form of a report. The unrounded minimum and maximum of $\varepsilon_\alpha$ for the zCDP guarantee specify the DP curves enclosing its power bound over the evaluation interval. For the final release, the power bounds derived from the zCDP guarantee apply to neighboring databases with identical invariants (Section~\ref{sec:census-scope}). Ranges are rounded to two decimal places; pointwise values are given in Table~\ref{tab:power-comparison}.}
\label{tab:report-example}
\centering
\small
\begin{tabular}{@{}p{0.35\linewidth}p{0.60\linewidth}@{}}
\toprule
Evaluation interval &
$0.01\le\alpha\le0.10$ \\
\addlinespace
Range of $\varepsilon_\alpha$ &
Approximately $2.93$--$4.25$; minimum at $\alpha\approx0.051$, maximum at $\alpha=0.01$ \\
\addlinespace
Catastrophic failure &
Excluded under this guarantee ($\varepsilon_0=0$) \\
\addlinespace
Original guarantee &
Noise-addition stage: $\rho$-zCDP with $\rho=2.631169245\ldots$ for persons and housing units, under replacement of one record \\
\addlinespace
Supplementary Gaussian approximation ($\mu=\sqrt{2\rho}$) &
$\varepsilon_\alpha$ approximately $2.13$--$3.89$, decreasing over the interval \\
\bottomrule
\end{tabular}
\end{table}

\subsection{Matching Guarantees in Utility Comparisons}\label{sec:reporting-mechanism-comparison}

Matching converted privacy parameters in a utility comparison can favor the mechanism whose power bound is loosened less by conversion. Section~\ref{sec:reporting-comparison} illustrates how reporting conversions introduce different amounts of slack for different mechanisms. Matching the converted values can therefore require adding more noise to the mechanism whose bound is loosened more. The resulting utility loss reflects both the mechanism's properties and the conversion used to match guarantees.

This issue arises in the NN-Wavelet and census tabulation studies discussed in Section~\ref{sec:introduction}, which match the Laplace mechanism's DP parameter $\varepsilon$ with the ADP $\varepsilon$ converted from the Gaussian mechanism's zCDP guarantee. Computing operationally equivalent $\varepsilon$ directly from the original guarantees at common significance levels would show how the power bounds differ under those comparison settings. Alternatively, the noise parameters could be adjusted to give the same $\varepsilon_\alpha$ at a specified $\alpha$, allowing utility to be compared under the same guaranteed power bound at that point. In both cases, operationally equivalent $\varepsilon$ expresses the bounds used for comparison as DP parameters giving the same constraints at the specified points, connecting the utility comparison to protection levels discussed in terms of DP's $\varepsilon$.

\subsection{Limitations and Considerations for Use}\label{sec:limitations}

Operationally equivalent $\varepsilon$ is only as tight as the input power bound. The mapping preserves that bound at the specified point, adding no conversion slack while retaining any slack already present. Even when mechanisms are matched using $\varepsilon_\alpha$ as in Section~\ref{sec:reporting-mechanism-comparison}, the gap between the common bound and the actual maximum power can differ across mechanisms. The gap depends on the tightness of the accounting or analysis used for each mechanism. Interpreting utility comparisons therefore requires considering how the input bounds were obtained.

We treat the Gaussian approximation in Section~\ref{sec:census-gaussian} as supplementary because the resulting curve has not been proved to bound test power for the composition of discrete Gaussian mechanisms used for P.L.~94-171. For the Demographic and Housing Characteristics File (DHC), \citet{SSW26} provide a non-asymptotic method for evaluating compositions of discrete Gaussian mechanisms with controlled numerical error. If this method can be applied to P.L.~94-171, reporting operationally equivalent $\varepsilon$ from the resulting power bounds would be preferable to using the Gaussian approximation. The Monte Carlo results of \citet[Section~6.3]{KAA+22} suggest that the resulting operationally equivalent $\varepsilon$ values would be close to those obtained from the Gaussian approximation.

Our Census reassessment applies under the adjacency relation and conditions on invariants specified in Section~\ref{sec:census-scope}. The published guarantee for noise addition also bounds test power after post-processing for neighboring databases with identical invariants. A broader assessment of protection must additionally consider what an adversary can distinguish from the information released exactly as invariants.

\section{Related Work}\label{sec:related-work}

\paragraph{Privacy accounting and operational evaluation of the 2020 U.S. Census}
\citet{KAA+22} interpret DP and its extensions from Bayesian and frequentist perspectives and apply these interpretations to the 2020 Census P.L.~94-171 redistricting data. They report power bounds at specified significance levels and Monte Carlo evaluations of compositions of discrete Gaussian mechanisms. The zCDP power bounds we compute agree with their reported values to two decimal places, and their Monte Carlo results support our supplementary Gaussian approximation (Sections~\ref{sec:census-zcdp} and~\ref{sec:census-gaussian}). Our reassessment verifies the published budgets and the calculation of ADP values, and uses operationally equivalent $\varepsilon$ to make explicit how ADP reporting changes the constraints conveyed by the original zCDP guarantee.

\citet{SSW25} use $f$-DP to analyze compositions of discrete Gaussian mechanisms for the 2020 Census DHC and obtain tighter privacy assessments than those based on the published guarantees. They use the mechanism's construction to tighten the analysis, whereas Section~\ref{sec:reporting-comparison} holds the input guarantee fixed to isolate the effect of the reporting format.

\paragraph{Reporting formats for privacy guarantees}
\citet{KH26} relate DP's $\varepsilon$ to GDP's $\mu$ using an adversary's ability to distinguish neighboring databases. Their calibration determines $\mu$ from $\varepsilon$ by matching the DP and GDP testing bounds at a specified significance level; solving the same condition for $\varepsilon$ gives the closed-form expression for $\varepsilon_\alpha(g_\mu^{\mathrm{GDP}})$ in Section~\ref{sec:operational-epsilon}. We share the aim of interpreting parameters across guarantee formats through their operational meaning. Our formulation takes general power bounds as input and supports both pointwise and interval reporting, with applications to zCDP guarantees and to comparisons between mechanisms.

\citet{GKK+26a} propose summarizing a privacy guarantee by a non-asymptotic GDP envelope and reporting its parameter $\mu$. This provides a single parameter describing a bound over all significance levels, but the envelope can loosen the input power bounds. Our approach preserves the input bound at each specified significance level and expresses it as the DP parameter giving the same constraint. Section~\ref{sec:reporting-slack} compares these reporting approaches.

\citet{SYSW26} show that, for a fixed pair of output distributions, the two directional privacy profiles and the two directional trade-off functions determine each other. They also quantify the information lost when a R\'enyi DP curve is summarized by a single zCDP parameter. Our comparisons hold the input guarantee fixed and examine how reporting conversions affect its power bounds at specified points. Direct reporting through operationally equivalent $\varepsilon$ preserves the original power bounds at those points.

\paragraph{Privacy estimation through attack experiments}
Privacy auditing uses observed false positive and false negative rates to obtain a statistical lower bound on a learning algorithm's privacy parameter $\varepsilon$. \citet{NHS+23} develop auditing methods that require few training runs. Epsilon* is an empirical privacy measure derived for a single trained model from the results of membership inference attacks \citep{NGA+24}. The closed-form expression for operationally equivalent $\varepsilon$ shares its two logarithmic terms with the $\delta=0$ case of the lower-bound estimator used in privacy auditing. Whereas auditing uses observed attack performance to assess how tight a privacy guarantee is, we apply the same expression to power bounds derived from guarantees, expressing their constraints at specified significance levels on the DP scale.

\section{Conclusion}\label{sec:conclusion}

We have presented a method for reporting general power-bound curves on a scale shared with DP\@. The $\alpha$-operationally equivalent $\varepsilon$ maps a mechanism's maximum power at significance level $\alpha$, or an upper bound on it derived from its guarantee, to the smallest DP parameter giving the same bound at that level. As a function of $\alpha$, it represents the input bound exactly on $(0,1)$, even when that bound is nonconcave. For power bounds corresponding to trade-off functions, and for mechanisms' maximum-power curves, the extrema of $\varepsilon_\alpha$ over an interval are determined by the endpoint values and the value at any switching point within the interval. When these extrema are finite, they specify the DP curves that enclose the input bound throughout the interval, providing a basis for concise interval reporting.

Our comparisons show that summarizing a guarantee by a single ADP pair with a small, fixed $\delta$ can substantially loosen its power bounds. For Laplace and Gaussian mechanisms with equal noise variance, single ADP pairs and GDP envelopes can also yield an ordering of the mechanisms opposite to that of their maximum-power curves at a specified significance level. Direct reporting with operationally equivalent $\varepsilon$ preserves the input bounds and their ordering at each specified point.

For the 2020 Census P.L.~94-171 redistricting data, expressing the constraints on distinguishing attacks in terms of DP's $\varepsilon$ clarifies the privacy protection that is difficult to discern from the large published value $\varepsilon=19.61$. The published zCDP guarantee yields operationally equivalent $\varepsilon$ values of approximately $4.25$, $2.94$, and $3.18$ at $\alpha=0.01$, $0.05$, and $0.10$, respectively. Converting the same guarantee to a single ADP pair with $\delta=10^{-10}$ gives power bounds close to $1$ at all three points. By comparing public documents with the implementation, we also reproduce $19.61$ as the sum of the rounded ADP values for persons and housing units. A supplementary Gaussian approximation of the noise-addition stage gives operationally equivalent $\varepsilon$ values of approximately $3.89$, $2.70$, and $2.13$ at the same three significance levels, respectively. In the debate over the large published value, these results provide a basis for assessing Census privacy protection in terms of the original guarantee's constraints on distinguishing attacks at specified significance levels.

For agencies reporting zCDP or GDP guarantees, the reporting format illustrated in Table~\ref{tab:report-example} provides a template: the range of $\varepsilon_\alpha$ over a stated interval of significance levels; the exclusion of catastrophic failure, indicated by $\varepsilon_0=0$; and the original guarantee with its accounting parameters. The original guarantee supports further privacy accounting, while operationally equivalent $\varepsilon$ expresses the constraints on distinguishing attacks on a scale shared with DP.

\section*{Acknowledgment}
This work was supported by JST K Program Grant Number JPMJKP24U5, Japan.

\section*{Code and Data Availability}
The Python scripts that reproduce all numerical results, tables, and figures, together with the pinned package versions and the expected output, are archived on Zenodo for long-term preservation \citep{TII+26-code} and are openly available under the MIT License. The scripts use no data files; the Census budgets are specified by the fractions listed in Appendix~\ref{app:tda-parameters}. The Census Bureau documents and DAS source code used in the analysis are available at the URLs cited in the references.

\bibliographystyle{abbrvnat-jpc}
\bibliography{op-equiv-jpc}

\appendix

\section{Lemmas and Proofs for Section~\ref{sec:operational-epsilon}}
\label{app:properties-proofs}

\begin{lem}[Ratio properties of maximum-power curves]\label{lem:maximum-power-ratios}
For every mechanism $M$, the functions
\[
\frac{G^M(\alpha)}{\alpha},\qquad
\frac{1-G^M(\alpha)}{1-\alpha}
\]
are nonincreasing on $(0,1)$. Moreover, $G^M$ is continuous on $(0,1)$.
\end{lem}

\begin{proof}
Let $i$ index the ordered neighboring pairs, and write $G_i$ for the corresponding maximum-power curves. Concavity of each $G_i$, together with $G_i(0)\ge0$ and $G_i(1)=1$, implies that $G_i(\alpha)/\alpha$ and $(1-G_i(\alpha))/(1-\alpha)$ are nonincreasing on $(0,1)$. We have
\[
\frac{G^M(\alpha)}{\alpha}=\sup_i\frac{G_i(\alpha)}{\alpha},\qquad
\frac{1-G^M(\alpha)}{1-\alpha}=\inf_i\frac{1-G_i(\alpha)}{1-\alpha}.
\]
Both ratios for $G^M$ are therefore nonincreasing, since pointwise suprema and infima preserve this monotonicity.

Since $G^M$ is nondecreasing, the monotonicity of $G^M(\alpha)/\alpha$ gives, for $0<s<t<1$,
\[
0\le G^M(t)-G^M(s)
\le\frac{t-s}{s}G^M(s)
\le\frac{t-s}{s}.
\]
Thus, $G^M$ is continuous on $(0,1)$.
\end{proof}

\begin{lem}[Recovery at zero]\label{lem:zero-endpoint-recovery}
For a mechanism $M$, let
\[
a_\infty
=\inf\left(
\{\alpha\in(0,1):\varepsilon_\alpha(G^M)=\infty\}
\cup\{1\}
\right).
\]
Then
\[
\lim_{\alpha\downarrow0}G^M(\alpha)=1-a_\infty.
\]
If $G^M$ is right-continuous at $0$, then $G^M(0)=1-a_\infty$.
\end{lem}

\begin{proof}
Put $r=\lim_{\alpha\downarrow0}G^M(\alpha)$, which exists because $G^M$ is nondecreasing. For $0<a<1$, the condition $\varepsilon_a(G^M)=\infty$ is equivalent to $G^M(a)=1$. If this condition holds, there is a sequence of tests, each for some ordered neighboring pair, with type I error probability at most $a$ and power tending to $1$. Reversing each pair and complementing the test gives tests with type I error probability tending to $0$ and power at least $1-a$. Hence $r\ge1-a$, and therefore $a_\infty\ge1-r$.

Conversely, if $r=0$, then $a_\infty\le1=1-r$. Otherwise, fix $a$ with $1-r<a<1$. Choose tests, each for some ordered neighboring pair, with type I error probability tending to $0$ and power tending to $r$. Reversing each pair and complementing the test gives tests with type I error probability eventually below $a$ and power tending to $1$. Thus, $G^M(a)=1$. Since this holds for every $a\in(1-r,1)$, we obtain $a_\infty\le1-r$. This proves the identity. Right-continuity at $0$ gives the final assertion.
\end{proof}

\begin{proof}[Proof of Proposition~\ref{prop:basic-properties}]
The first identity in part~(i) follows by substituting $g=g_\varepsilon^{\mathrm{DP}}$ into~\eqref{eq:operational-epsilon}.

For the second identity, put
\[
\varepsilon_{\mathrm{sup}}^M
=\sup_{0<\alpha<1}\varepsilon_\alpha(G^M).
\]
For any finite $\varepsilon\ge0$,
equations~\eqref{eq:operational-epsilon}
and~\eqref{eq:pure-dp-power} give
\[
\begin{aligned}
\varepsilon_{\mathrm{sup}}^M\le\varepsilon
&\iff \varepsilon_\alpha(G^M)\le\varepsilon
&&\text{for all }0<\alpha<1\\
&\iff G^M(\alpha)\le g_\varepsilon^{\mathrm{DP}}(\alpha)
&&\text{for all }0<\alpha<1.
\end{aligned}
\]

Since $G^M$ is nondecreasing, the last condition implies
\[
0\le G^M(0)\le G^M(\alpha)\le e^\varepsilon\alpha
\qquad(0<\alpha<1).
\]
Letting $\alpha\downarrow0$ gives $G^M(0)=0$.
Since both curves equal $1$ at $\alpha=1$, the bound extends
to $[0,1]$.

Consequently, if $\varepsilon_{\mathrm{sup}}^M<\infty$,
then $M$ satisfies $\varepsilon$-DP if and only if
$\varepsilon\ge\varepsilon_{\mathrm{sup}}^M$;
if $\varepsilon_{\mathrm{sup}}^M=\infty$, no finite DP
parameter exists.
Taking the infimum over the finite DP parameters for $M$
proves the second identity, with $\inf\emptyset=\infty$.

For part~(ii), fix $0<\alpha<1$. The function $\varepsilon\mapsto g_\varepsilon^{\mathrm{DP}}(\alpha)$ is strictly increasing and continuous, equals $\alpha$ at $\varepsilon=0$, and tends to $1$ as $\varepsilon\to\infty$. The stated bijection then follows from Definition~\ref{def:operational-epsilon}. Substituting~\eqref{eq:operational-epsilon} into~\eqref{eq:pure-dp-power} gives the inverse formula when $\varepsilon_\alpha(g)$ is finite. If $\varepsilon_\alpha(g)=\infty$, then~\eqref{eq:operational-epsilon} gives $g(\alpha)=1$. At $\alpha=0$, finiteness of $\varepsilon_0(g)$ implies $g(0)=0$ and $\varepsilon_0(g)=0$; at $\alpha=1$, we have $g(1)=1$ and $\varepsilon_1(g)=0$. The equality of bounds therefore also holds at both endpoints whenever $\varepsilon_\alpha(g)$ is finite.

For part~(iii), the guarantee for $M$ and the assumption on $g$ give
\[
G^M(\alpha_0)\le g(\alpha_0)=\alpha_0.
\]
Suppose that $M$ does not satisfy $0$-DP\@. By the equivalence in Section~\ref{sec:hypothesis-testing}, some neighboring database pair admits a test $\phi_1$ whose type I error probability $\alpha_1$ and power $\pi_1$ satisfy $\pi_1>\alpha_1$.

If $\alpha_0<\alpha_1$, let $\theta=\alpha_0/\alpha_1$ and randomize between $\phi_1$ and the test that never rejects, choosing $\phi_1$ with probability $\theta$. The resulting test has type I error probability $\alpha_0$ and power $\theta\pi_1>\theta\alpha_1=\alpha_0$. If $\alpha_0\ge\alpha_1$, let $\theta=(1-\alpha_0)/(1-\alpha_1)$ and randomize between $\phi_1$ and the test that always rejects, choosing $\phi_1$ with probability $\theta$. This test also has type I error probability $\alpha_0$, and its power exceeds $\alpha_0$ by $\theta(\pi_1-\alpha_1)>0$. In either case, $G^M(\alpha_0)>\alpha_0$, contradicting $G^M(\alpha_0)\le g(\alpha_0)=\alpha_0$. Hence $M$ satisfies $0$-DP.

Part~(iv) follows because every $\varepsilon$ satisfying $g_\varepsilon^{\mathrm{DP}}(\alpha)\ge g_2(\alpha)$ in Definition~\ref{def:operational-epsilon} also satisfies $g_\varepsilon^{\mathrm{DP}}(\alpha)\ge g_1(\alpha)$.

For part~(v), $g$ is continuous on $(0,1)$, and both $g(\alpha)/\alpha$ and $(1-g(\alpha))/(1-\alpha)$ are nonincreasing there. For $g=1-f$, these properties follow from concavity of $g$, together with $g(0)\ge0$ and $g(1)=1$. For $g=G^M$, they follow from Lemma~\ref{lem:maximum-power-ratios}.

The monotonicity of these ratios implies that the first logarithmic term in~\eqref{eq:operational-epsilon} is nonincreasing and the second is nondecreasing on $(0,1)$, with the second interpreted as $\infty$ when $g(\alpha)=1$. For $\alpha\in[a,b]$, the first term is bounded above by its value at $a$, and the second by its value at $b$. Thus, $\varepsilon_\alpha(g)\le\max\{\varepsilon_a(g),\varepsilon_b(g)\}$ throughout $[a,b]$, proving the assertion about the maximum.

For the minimum, the two logarithmic terms coincide at a switching point and are monotone in opposite directions, so $\varepsilon_\alpha(g)$ is minimized there. Substitution gives the stated value. If no switching point lies in $[a,b]$, continuity of $g$ ensures that the same logarithmic term dominates throughout the interval. Its monotonicity then implies that the minimum is attained at an endpoint.

For part~(vi), a test based on the post-processed output can also be performed on the original output, so $G^{A\circ M}(\alpha)\le G^M(\alpha)$. The result follows from part~(iv).
\end{proof}

\section{Slack in \texorpdfstring{$f$}{f}-DP and Privacy Profile Bounds for a Nonconcave Maximum-Power Curve}
\label{app:nonconcave-power}

The mechanism in the following example has a nonconcave maximum-power curve. Its tightest $f$-DP guarantee and full privacy profile yield the same power bound, which strictly exceeds the maximum-power curve at some significance levels. It also illustrates the application of Proposition~\ref{prop:basic-properties}(v) without concavity.

Let $M$ have input and output spaces $\{0,1\}$, with the two inputs treated as neighboring databases, and let $P=M(0)=\operatorname{Ber}(1/4)$ and $Q=M(1)=\operatorname{Ber}(1/2)$.

The most powerful test assigns rejection probability to output $1$ before output $0$ when $P$ is the null distribution, and in the reverse order when $Q$ is the null distribution. Randomization at the threshold gives
\[
G^{P,Q}(\alpha)
=\min\left\{2\alpha,\frac{1+2\alpha}{3}\right\},
\qquad
G^{Q,P}(\alpha)
=\min\left\{\frac{3\alpha}{2},\frac{1+\alpha}{2}\right\}.
\]
Taking the maximum over the two directions yields
\[
G^M(\alpha)=
\begin{cases}
2\alpha, & 0\le\alpha\le1/4,\\
(1+2\alpha)/3, & 1/4<\alpha\le2/5,\\
3\alpha/2, & 2/5<\alpha\le1/2,\\
(1+\alpha)/2, & 1/2<\alpha\le1.
\end{cases}
\]
This curve is not concave: its slope increases from $2/3$ to $3/2$ at $\alpha=2/5$.

The greatest convex function below $1-G^M$ is
\[
f(\alpha)
=\max\left\{
1-2\alpha,\;
\frac34-\alpha,\;
\frac{1-\alpha}{2}
\right\}.
\]
The function $f$ is continuous, convex, and nonincreasing,
with $f(\alpha)\le1-\alpha$ and $f=f^{-1}$.
It is therefore a symmetric trade-off function
\citep[Proposition~1]{DRS22} and gives the tightest
$f$-DP guarantee for $M$.

The power bound corresponding to this $f$-DP guarantee is
\[
g^f(\alpha)
=1-f(\alpha)
=\min\left\{
2\alpha,\alpha+\frac14,\frac{1+\alpha}{2}
\right\}.
\]
This bound coincides with $G^M$ on
$[0,1/4]\cup[1/2,1]$ and strictly exceeds it on $(1/4,1/2)$.
For example,
\[
G^M(3/8)=\frac7{12}<\frac58=g^f(3/8).
\]

For $\varepsilon\ge0$, let $\delta_M(\varepsilon)$ be the
smallest $\delta$ such that $M$ satisfies
$(\varepsilon,\delta)$-ADP\@.
Since $1-g_{\varepsilon,\delta}^{\mathrm{ADP}}$ is convex,
the maximality of $f$ ensures that the ADP guarantees
implied by $f$-DP are exactly those satisfied by $M$.
The duality in \citet[Proposition~6]{DRS22} therefore gives
\[
\begin{aligned}
\delta_M(\varepsilon)
&=\sup_{0\le\alpha\le1}
\left\{1-f(\alpha)-e^\varepsilon\alpha\right\}\\
&=\max\left\{
0,\frac12-\frac{e^\varepsilon}{4},
\frac34-\frac{e^\varepsilon}{2},
1-e^\varepsilon
\right\}\\
&=\max\left\{0,\frac12-\frac{e^\varepsilon}{4}\right\}.
\end{aligned}
\]
The supremum of the piecewise linear expression is attained
at an endpoint or breakpoint,
$\alpha\in\{0,1/4,1/2,1\}$, giving the second equality.
The last equality uses $e^\varepsilon\ge1$.

By the same duality, the full privacy profile yields
this same power bound:
\[
\inf_{\varepsilon\ge0}
g_{\varepsilon,\delta_M(\varepsilon)}^{\mathrm{ADP}}(\alpha)
=g^f(\alpha).
\]

\begin{figure}[tbp]
\centering
\includegraphics[width=\linewidth]{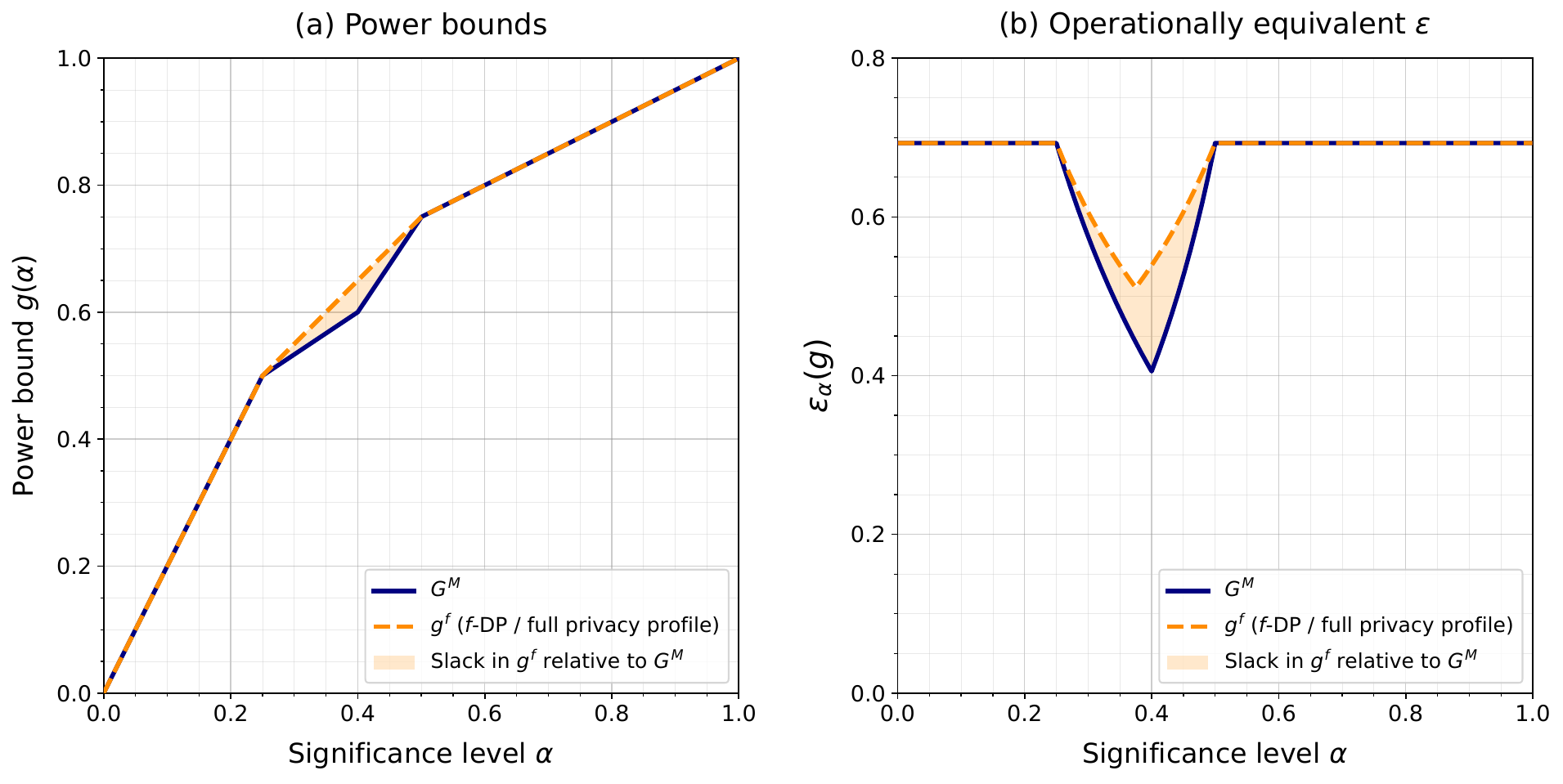}
\caption{The nonconcave maximum-power curve $G^M$ and the common bound $g^f$ from the tightest $f$-DP guarantee and the full privacy profile, shown on the power (a) and operationally equivalent $\varepsilon$ (b) scales. Shading indicates the slack in $g^f$ relative to $G^M$. Panel (b) omits the endpoint values $\varepsilon_0=\varepsilon_1=0$.}
\label{fig:nonconcave-power}
\end{figure}

Thus, the full privacy profile has the same slack relative to $G^M$
as the tightest $f$-DP guarantee.
Figure~\ref{fig:nonconcave-power} shows that this slack is confined
to $(1/4,1/2)$ on both the power and operationally equivalent
$\varepsilon$ scales.
Direct reporting through $\varepsilon_\alpha(G^M)$ retains
$G^M$ itself on $(0,1)$ by
Proposition~\ref{prop:basic-properties}(ii).

Proposition~\ref{prop:basic-properties}(v) applies to this nonconcave $G^M$. Its switching point is $\alpha_*=2/5$, where $G^M(\alpha_*)=3/5=1-\alpha_*$. Hence
\[
\min_{0<\alpha<1}\varepsilon_\alpha(G^M)
=\log\frac32.
\]

\section{Reporting Comparisons for the Laplace and Gaussian Mechanisms}
\label{app:reporting-parameters}

This appendix gives the calculations underlying Section~\ref{sec:reporting-slack}. Both mechanisms add noise to the same scalar query of sensitivity $1$. We assume that the query values differ by $1$ for at least one pair of neighboring databases. The Laplace scale is $1$, and the Gaussian standard deviation is $\sqrt{2}$, so both noise variances are $2$.

\subsection{Maximum-Power Curves}\label{app:laplace-gaussian-computation}

The Laplace mechanism's maximum power in Table~\ref{tab:laplace-gaussian-comparison} is obtained by testing $P=\operatorname{Lap}(0,1)$ against $Q=\operatorname{Lap}(1,1)$, where $\operatorname{Lap}(m,b)$ denotes the Laplace distribution with location $m$ and scale $b$.
Let $X$ denote the output random variable.
For a possible output $x\in\mathbb R$, the log-likelihood ratio is
\[
\log\frac{dQ}{dP}(x)=\lvert x\rvert-\lvert x-1\rvert,
\]
which is nondecreasing in $x$.
For $0<\alpha<1$, a threshold test with
$P(X\ge t)=\alpha$ therefore attains maximum power.
With $\pi=Q(X\ge t)$, the Laplace tail probabilities give
\[
\begin{aligned}
t\ge1:\quad
&\frac{\pi}{\alpha}
=\frac{e^{1-t}/2}{e^{-t}/2}
=e,\\[4pt]
0<t<1:\quad
&\alpha(1-\pi)
=\frac{e^{-t}}2\frac{e^{t-1}}2
=\frac{e^{-1}}4,\\[4pt]
t\le0:\quad
&\frac{1-\pi}{1-\alpha}
=\frac{e^{t-1}/2}{e^t/2}
=e^{-1}.
\end{aligned}
\]
The thresholds $t=1$ and $t=0$ correspond to
$\alpha=e^{-1}/2$ and $\alpha=1/2$, respectively.
Solving these relations for $\pi$ yields
\begin{equation}
G^{\mathrm{Lap}}(\alpha)=
\begin{cases}
e\alpha, & 0\le\alpha\le e^{-1}/2,\\
1-\dfrac{e^{-1}}{4\alpha}, & e^{-1}/2<\alpha<1/2,\\
1-e^{-1}(1-\alpha), & 1/2\le\alpha\le1.
\end{cases}
\label{eq:laplace-power-example}
\end{equation}
The endpoints $\alpha=0$ and $\alpha=1$ correspond to
the tests that never reject and always reject, respectively.

Reversing the null and alternative distributions gives the same curve by reflection symmetry. For a location difference $d\in[0,1]$, replace $e$ and $e^{-1}$ in~\eqref{eq:laplace-power-example} by $e^d$ and $e^{-d}$. Maximum power increases with $d$. Thus, for a mechanism with sensitivity $1$ and a neighboring pair attaining that difference, the maximum power is given by~\eqref{eq:laplace-power-example}.

As noted in Section~\ref{sec:gaussian-mechanism},
the Gaussian mechanism's maximum-power curve equals
$g_{\Delta/\sigma}^{\mathrm{GDP}}$.
With $\Delta=1$ and $\sigma=\sqrt{2}$, we therefore have
\[
G^{\mathrm{Gau}}(\alpha)
=g_{1/\sqrt{2}}^{\mathrm{GDP}}(\alpha).
\]
Substituting both curves into~\eqref{eq:operational-epsilon}
gives the values in Table~\ref{tab:laplace-gaussian-comparison}.

\subsection{Single ADP Pairs}\label{sec:adp-point-selection}\label{app:adp-tangent-pairs}

We consider two ways of selecting ADP pairs: fixing a small $\delta$, and choosing a pair separately for each evaluation point so that its power bound is tangent to the mechanism's maximum-power curve there.

\paragraph{Pairs with a fixed small $\delta$}
We retain the Laplace mechanism's $1$-DP guarantee as the pair $(1,0)$. For the Gaussian mechanism, we follow the convention of fixing a small $\delta$ described in Section~\ref{sec:adp-reporting} and consider
\[
\delta\in\{10^{-6},10^{-8},10^{-10},10^{-12},10^{-14}\}.
\]
For each value of $\delta$, we solve~\eqref{eq:gdp-adp-profile} for $\varepsilon$ with $\mu=1/\sqrt{2}$. We denote the resulting ADP parameter by $\varepsilon^{\mathrm{ADP}}(\delta)$.
Substituting these pairs into~\eqref{eq:adp-power} gives the power bounds at each evaluation point.

The values reported for the Gaussian mechanism vary substantially even over this range of small $\delta$ values. As Table~\ref{tab:adp-mechanism-reporting} shows, reducing $\delta$ from $10^{-6}$ to $10^{-14}$ increases $\varepsilon^{\mathrm{ADP}}(\delta)$ from approximately $3.31$ to $5.43$. At $\alpha=0.01$, for example, the power bound conveyed by the ADP pair changes from approximately $0.2732$ to $0.9957$. The mechanism's maximum power $G^{\mathrm{Gau}}(0.01)\approx0.0527$ and its operationally equivalent $\varepsilon_{0.01}(G^{\mathrm{Gau}})\approx1.6620$ remain unchanged. Thus, the magnitude of the reported $\varepsilon$ and the constraint on distinguishability conveyed by the report depend on the choice of $\delta$ as well as on the mechanism itself.

For all five choices of $\delta$ in Table~\ref{tab:adp-mechanism-reporting}, the ADP reports give a smaller power bound for the Laplace mechanism at all five evaluation points. This ordering reverses the maximum-power comparison at $\alpha=0.25$ and $0.50$ in Table~\ref{tab:laplace-gaussian-comparison}. The Gaussian ADP bound is loose at all five points for every choice of $\delta$, whereas the Laplace $1$-DP bound is exact at four of these points; at $\alpha=0.25$, it gives $0.6796$ compared with the maximum power $0.6321$.

\begin{table}[t]
\caption{ADP conversions and power bounds for the Gaussian mechanism in Table~\ref{tab:laplace-gaussian-comparison}, for different choices of $\delta$. Values are rounded.}
\label{tab:adp-mechanism-reporting}
\centering
\small
\begin{tabular}{@{}rrrrrrr@{}}
\toprule
& & \multicolumn{5}{c}{Power bound implied by the ADP pair} \\
\cmidrule(l){3-7}
$\delta$ & $\varepsilon^{\mathrm{ADP}}(\delta)$ & $\alpha=0.01$ & $0.05$ & $0.10$ & $0.25$ & $0.50$ \\
\midrule
$10^{-6}$ & 3.3076 & 0.273196 & 0.965226 & 0.967057 & 0.972547 & 0.981698 \\
$10^{-8}$ & 3.9425 & 0.515455 & 0.981570 & 0.982540 & 0.985450 & 0.990300 \\
$10^{-10}$ & 4.4922 & 0.893194 & 0.989364 & 0.989924 & 0.991603 & 0.994402 \\
$10^{-12}$ & 4.9835 & 0.993218 & 0.993492 & 0.993835 & 0.994862 & 0.996575 \\
$10^{-14}$ & 5.4314 & 0.995667 & 0.995842 & 0.996061 & 0.996717 & 0.997812 \\
\bottomrule
\end{tabular}
\end{table}

\paragraph{Pairs tangent at the specified point}
Selecting an ADP pair separately for each evaluation point can remove the slack there, but may require a $\delta$ that cannot be treated as negligible. Table~\ref{tab:adp-tangent-pairs} gives such pairs for both mechanisms.

\begin{table}[t]
\caption{ADP pairs $(\varepsilon^{\mathrm t},\delta^{\mathrm t})$ whose power bounds are tangent to the maximum-power curves of the mechanisms in Table~\ref{tab:laplace-gaussian-comparison} at the specified significance level. Values are rounded.}
\label{tab:adp-tangent-pairs}
\centering
\small
\begin{tabular}{@{}rrrrr@{}}
\toprule
& \multicolumn{2}{c}{Laplace} & \multicolumn{2}{c}{Gaussian} \\
\cmidrule(lr){2-3}\cmidrule(lr){4-5}
$\alpha$ & $\varepsilon^{\mathrm t}$ & $\delta^{\mathrm t}$ & $\varepsilon^{\mathrm t}$ & $\delta^{\mathrm t}$ \\
\midrule
0.01 & 1.000000 & 0 & 1.394976 & 0.012349 \\
0.05 & 1.000000 & 0 & 0.913087 & 0.049587 \\
0.10 & 1.000000 & 0 & 0.656194 & 0.090089 \\
0.25 & 0.386294 & 0.264241 & 0.226936 & 0.199323 \\
0.50 & 1.000000 & 0 & 0.250000 & 0.192155 \\
\bottomrule
\end{tabular}
\end{table}

For the Laplace mechanism, the power bound from $(1,0)$ already matches the maximum-power curve at four of the five evaluation points; at $\alpha=0.25$, a tangent pair is
\[
(\varepsilon^{\mathrm t},\delta^{\mathrm t})
=(\log 4-1,\ 1-2/e).
\]

This pair is obtained from the tangent to the curve in~\eqref{eq:laplace-power-example} at $\alpha=1/4$. At this point,
\[
G^{\mathrm{Lap}}(1/4)=1-\frac1e,
\qquad
(G^{\mathrm{Lap}})'(1/4)=\frac4e,
\]
so the tangent line is
\[
\begin{aligned}
\pi
&=\left(1-\frac1e\right)+\frac4e\left(\alpha-\frac14\right)\\
&=\frac4e\alpha+1-\frac2e.
\end{aligned}
\]
Matching its slope and intercept with the ADP branch $e^\varepsilon\alpha+\delta$ gives the pair above. The other linear branch in~\eqref{eq:adp-power} becomes $1/2+(e/4)\alpha$, which is tangent to the same curve at $\alpha=1/e$. Since $G^{\mathrm{Lap}}$ is concave, both tangent lines bound it from above throughout $[0,1]$. Their minimum with $1$ therefore gives a valid ADP bound for the mechanism, with equality at $\alpha=1/4$.

The tangent ADP pairs for the Gaussian mechanism follow from the GDP-to-ADP conversion of \citet[Corollary~1 and its proof in Appendix~A]{DRS22}. We set $\mu=1/\sqrt{2}$ and solve the contact-point condition for $\varepsilon$. Accounting for the two symmetric contact points gives, for $0<\alpha<1$,
\[
\varepsilon^{\mathrm t}(\alpha)
=\left\lvert\frac{\Phi^{-1}(\alpha)}{\sqrt{2}}+\frac14\right\rvert,
\qquad
\delta^{\mathrm t}(\alpha)
=\delta_{1/\sqrt{2}}\!\left(\varepsilon^{\mathrm t}(\alpha)\right).
\]
Each pair reproduces its mechanism's maximum power at the specified point.

\subsection{GDP Envelopes}\label{sec:gdp-envelope-comparison}\label{app:gdp-envelope}

For each mechanism $M$, we determine the smallest $\mu$ satisfying
\[
G^M(\alpha)\le g_\mu^{\mathrm{GDP}}(\alpha)
\qquad\text{for every }\alpha\in[0,1].
\]

The Gaussian curve is its own envelope, with
\[
\mu_{\mathrm{Gau}}=\frac1{\sqrt2}\approx0.7071.
\]

The minimum GDP envelope parameter for the Laplace curve is
\begin{equation}
\mu_{\mathrm{Lap}}
=-2\Phi^{-1}(a_0)
=1.030063997\ldots,
\qquad a_0=\frac{e^{-1/2}}2.
\label{eq:laplace-gdp-minimum-mu}
\end{equation}
We first show that any GDP envelope requires $\mu\ge\mu_{\mathrm{Lap}}$, then verify that $\mu=\mu_{\mathrm{Lap}}$ gives a valid envelope on $[0,1]$.

The point $a_0$ lies in the middle interval of~\eqref{eq:laplace-power-example} and satisfies $G^{\mathrm{Lap}}(a_0)=1-a_0$. Any GDP envelope with parameter $\mu$ must therefore satisfy
\[
\Phi\bigl(\Phi^{-1}(a_0)+\mu\bigr)
\ge1-a_0=\Phi\bigl(-\Phi^{-1}(a_0)\bigr).
\]
Monotonicity of $\Phi$ gives $\mu\ge-2\Phi^{-1}(a_0)=\mu_{\mathrm{Lap}}$.

To show sufficiency, take $\alpha\in[e^{-1}/2,1/2]$ and put $\beta=1-G^{\mathrm{Lap}}(\alpha)=e^{-1}/(4\alpha)$. Then $\alpha\beta=a_0^2$. Log-concavity of the normal distribution function gives
\[
\Phi\!\left(\frac{\Phi^{-1}(\alpha)+\Phi^{-1}(\beta)}2\right)
\ge\sqrt{\alpha\beta}=a_0.
\]
Consequently, $\Phi^{-1}(\alpha)+\Phi^{-1}(\beta)\ge2\Phi^{-1}(a_0)=-\mu_{\mathrm{Lap}}$, which yields
\[
G^{\mathrm{Lap}}(\alpha)=1-\beta
\le\Phi\bigl(\Phi^{-1}(\alpha)+\mu_{\mathrm{Lap}}\bigr)
=g_{\mu_{\mathrm{Lap}}}^{\mathrm{GDP}}(\alpha).
\]
The Laplace curve is linear on each of the remaining intervals, $[0,e^{-1}/2]$ and $[1/2,1]$. On each interval, the GDP curve is concave and lies at or above the Laplace curve at both endpoints, so the inequality holds throughout the interval. Thus, the parameter in~\eqref{eq:laplace-gdp-minimum-mu} defines a valid envelope over $[0,1]$ and is the smallest possible. Both curves have derivative $1$ at $a_0$, where they touch.

\section{TDA zCDP Budgets and the Gaussian Approximation Parameter}
\label{app:tda-parameters}

The Certificates of Disclosure Avoidance for P.L.~94-171 US Persons and US Units are published as separate files \citep{USCB21e}. Page~5 of each PDF records the corresponding zCDP parameter:
\[
\begin{aligned}
\rho_{\mathrm p}
&=\frac{293764}{114921}
=2.556225581\ldots,\\
\rho_{\mathrm h}
&=\frac{11417641}{152349649}
=0.074943664\ldots.
\end{aligned}
\]
Composition gives
\[
\begin{aligned}
\rho
&=\rho_{\mathrm p}+\rho_{\mathrm h}\\
&=\frac{293764}{114921}
+\frac{11417641}{152349649}\\
&=2.631169245\ldots.
\end{aligned}
\]
Rounded to two decimal places, these values agree with the published $2.56$ for persons, $0.07$ for housing units, and $2.63$ after composition \citep{USCB21f}.

Section~\ref{sec:census-gaussian} approximates TDA's noise-addition stage by a composition of the corresponding continuous Gaussian mechanisms. Their GDP guarantees and the GDP composition rule give $\mu=\sqrt{2\rho}=2.293978746\ldots$ for this approximation.

\end{document}